\documentclass[aps,prd,nofootinbib,onecolumn,showpacs,superscriptaddress,
preprintnumbers,floatfix,notitlepage]{revtex4-1}

\usepackage{graphicx}
\usepackage{graphics}
\usepackage{amsmath}
\usepackage{bm}% bold math

\newcommand{\hermes}{\textsc{Hermes }}
\newcommand{\compass}{\textsc{Compass }}
\newcommand{\minuit}{\textsc{Minuit }}

\newcommand{\xbj}{x}

\newcommand{\smarrow}{\mbox{\raisebox{-4.5pt}[0pt][0pt]{$\hspace{-1pt}
      \vec{\phantom{v}}$}}}
\newcommand{\T}{\perp}
\newcommand{\Tperp}{T}

\begin{document}

\allowdisplaybreaks[2]

%\preprint{}

\title{Investigations into the flavor dependence of partonic transverse momentum}


\author{Andrea Signori}
\email{asignori@nikhef.nl}
\affiliation{Nikhef Theory Group and 
Department of Physics and Astronomy, VU University Amsterdam \\
De Boelelaan 1081, NL-1081 HV Amsterdam, the Netherlands}

\author{Alessandro Bacchetta}
\email{alessandro.bacchetta@unipv.it}
\affiliation{Dipartimento di Fisica, Universit\`a di Pavia}
\affiliation{INFN Sezione di Pavia, via Bassi 6, 27100 Pavia, Italy}

\author{Marco Radici}
\email{marco.radici@pv.infn.it}
\affiliation{INFN Sezione di Pavia, via Bassi 6, 27100 Pavia, Italy}

\author{Gunar Schnell}
\email{gunar.schnell@ehu.es}
\affiliation{Department of Theoretical Physics, University of the Basque Country UPV/EHU, 48080 Bilbao, 
Spain}
\affiliation{IKERBASQUE, Basque Foundation for Science, 48011 Bilbao, Spain}

%\runninglinenumbers

\preprint{NIKHEF 2013-030}

\begin{abstract}
Recent experimental data on semi-inclusive deep-inelastic scattering from
the \hermes  collaboration allow us to discuss for the first time the flavor
dependence of unpolarized transverse-momentum dependent distribution and
fragmentation functions.  
We find convincing indications that favored fragmentation functions into pions
have smaller average transverse momentum than
unfavored functions and
fragmentation functions into kaons. 
We find weaker indications of flavor dependence in the distribution functions.
\end{abstract}

%\keywords{}

\maketitle

%%%%%%%%%%%%%%%%%%%%%%%%%%%%%%%%%%%%%%%%%%%%
\section{Introduction}
%%%%%%%%%%%%%%%%%%%%%%%%%%%%%%%%%%%%%%%%%%%%

Transverse-momentum-dependent (TMD) parton distribution functions (PDFs) and
fragmentation functions (FFs)
give a multi-dimensional description of partonic structure in momentum space. 
They are functions of the longitudinal and transverse momentum of partons, with
respect to the reference hadron momentum.  
As such, they offer richer information compared to standard
collinear
PDFs and FFs, which depend only
on the longitudinal momentum. 
In the last decade, TMD PDFs and FFs have gained increasing attention
especially because of emerging data from experiments on semi-inclusive
deep-inelastic scattering (DIS) (for reviews see, e.g.,
\cite{Barone:2010zz,Boer:2011fh,Aidala:2012mv}).   

In spite of this progress, we have still little knowledge about the most
simple and most common of all TMD PDFs: 
the ``unpolarized'' distribution, $f_1^a(x, \bm{k}_\T^2)$, i.e., the
distribution of partons with flavor $a$ summed over their polarization and
averaged over the polarization of the parent hadron.  
The features of the corresponding collinear standard PDF $f_1^a(x)$ strongly
depend on the parton flavor $a$ (see, e.g., Refs.~\cite{Forte:2013wc,%
Gao:2013xoa,Owens:2012bv,Ball:2012cx,Martin:2009iq,JimenezDelgado:2008hf}). It
comes natural, 
therefore, to question whether or not partons of different flavors have
different transverse-momentum distributions. 
Several model calculations predict different transverse-momentum behaviors for
different 
quarks~\cite{Bacchetta:2008af,Bacchetta:2010si,Wakamatsu:2009fn,%
Efremov:2010mt,Bourrely:2010ng,Matevosyan:2011vj,Schweitzer:2012hh},
although others do not~\cite{Pasquini:2008ax,Lorce:2011dv,Avakian:2010br}. 
Indications of flavor dependence in TMD PDFs come also from pioneering studies in
lattice QCD~\cite{Musch:2010ka}. Therefore, we believe there are compelling
motivations to study the flavor dependence of TMD PDFs. 

The measurements recently published by the \hermes
collaboration~\cite{Airapetian:2012ki} are ideal to 
address this issue, since they refer to semi-inclusive DIS off different targets (protons
and deuterons), with different final-state hadrons (charge-separated pions 
and kaons), and with multidimensional binning. This is a landmark achievement
in the knowledge of the internal structure of hadrons. Earlier data already
gave some indications, but were limited in the variety of targets, or
final-state hadrons, or multidimensional coverage (see, e.g.,
\cite{Arneodo:1984rm,Adloff:1996dy,Mkrtchyan:2007sr,%
Osipenko:2008aa,Asaturyan:2011mq}).   

The \compass collaboration has recently released similar
data~\cite{Adolph:2013stb}.  The amount of
statistics is in this case impressive and the kinematic coverage is in general
wider than at {\sc Hermes}. However, at the moment these data are available
only for deuteron targets and for unidentified final charged
hadrons. Therefore, we decided not to use these data, although they will
certainly play an essential role in the near future. 

Dealing with semi-inclusive DIS, we need to consider also fragmentation functions and their
transverse-momentum dependence. Also in this case, it is possible that
different quark flavors fragment into different hadrons with characteristic
transverse-momentum 
distributions~\cite{Bacchetta:2007wc,Matevosyan:2011vj}. 
This is another
fundamental  question that has
never been addressed at the phenomenological level.  

Since our work represents one of the first explorations on this topic, we
adopt here a simplified framework, essentially based on a parton-model
picture. We perform a leading-order analysis and neglect any
modification 
that can be induced by QCD
evolution, both in the collinear PDFs and FFs as well as in the TMD ones. 
This approximation is justified by the limited range in $Q^2$ of
data: no difficulty arises in describing them with this simplified
framework. All our assumptions, the notation, and the general theoretical
framework are briefly outlined in Sec.~\ref{s:theory}. In
Sec.~\ref{s:analysis}, we describe our fitting procedure.  
In Sec.~\ref{s:results}, we present our final results, and in
Sec.~\ref{s:conclusion} we draw some conclusions and outlooks.

%%%%%%%%%%%%%%%%%%%%%%%%%%%%%%%%%%%%%%%%%%%%
\section{Theoretical framework}
%%%%%%%%%%%%%%%%%%%%%%%%%%%%%%%%%%%%%%%%%%%%
% better to introduce subsections
\label{s:theory}

In one-particle semi-inclusive DIS, a lepton $\ell$ with momentum $l$ scatters to a final
state with momentum $l'$ off a hadron target $N$ with mass $M$ and momentum
$P$, producing (at least) one hadron $h$ in the final state with mass $M_h$
and momentum $P_h$:  
\begin{equation}
  \label{e:sidis}
\ell(l) + N(P) \to \ell(l') + h(P_h) + X \, .
\end{equation}
The space-like momentum transfer is $q = l - l'$, with $Q^2 = - q^2$. We
introduce the usual invariants  
\begin{align}
  \label{e:xyz}
\xbj &= \frac{Q^2}{2\,P\cdot q},
&
y &= \frac{P \cdot q}{P \cdot l},
&
z &= \frac{P \cdot P_h}{P\cdot q},
&
\gamma &= \frac{2 M x}{Q} .
\end{align}

The available data refer to hadron multiplicities in semi-inclusive DIS, namely to the differential number of hadrons produced per corresponding inclusive DIS event. In terms of cross sections, we define the multiplicities as
\begin{equation}
m_N^h (x,z,\bm{P}_{h\Tperp}^2, Q^2) = \frac{d \sigma_N^h / dx  dz d\bm{P}_{h\Tperp}^2 dQ^2}
                                                                   {d\sigma_{\text{DIS}} / dx dQ^2 }\, ,
\label{e:multiplicity}
\end{equation}
where $d\sigma_N^h$ is the differential cross section for the semi-inclusive DIS process and $d\sigma_{\text{DIS}}$ is the corresponding inclusive one, 
and where \( \bm{P}_{h\Tperp} \) is the component of \( \bm{P}_{h} \) transverse to \( \bm{q} \). 
In the single-photon-exchange approximation, the multiplicities can be written as ratios of
structure functions (see \cite{Bacchetta:2006tn} for details):
\begin{equation}
m_N^h (x,z,\bm{P}_{h\Tperp}^2, Q^2) =   
\frac{ \pi\, F_{UU ,T}(x,z,\bm{P}_{h\Tperp}^2, Q^2) + \pi \, \varepsilon  F_{UU ,L}(x,z,\bm{P}_{h\Tperp}^2, Q^2)}
        {F_{T}(x,Q^2) + \varepsilon  F_{L}(x,Q^2)} \, ,
 \label{e:mFF}
\end{equation} 
where
\begin{align}
\varepsilon &= \frac{1-y -\frac{1}{4} \gamma^2 y^2}{1-y+\frac{1}{2} y^2 +\frac{1}{4} \gamma^2 y^2} \, .
\end{align}  
We recall that the notation $F_{XY,Z}$ indicates the response of the hadron target with polarization $Y$ to a lepton beam with polarization $X$ and for the virtual photon exchanged in the polarization state $Z$. Therefore, the numerator of Eq.~\eqref{e:mFF} involves semi-inclusive DIS processes with only unpolarized beam and target. We remark that the above expressions assume a complete integration over the azimuthal angle of the detected hadron. Acceptance effects may modify these formulae, due to the presence of azimuthal modulations in the cross section, though for the data used here such effects were included in the systematic uncertainties.

We consider the limits $M^2/Q^2 \ll 1$ and $\bm{P}_{h\Tperp}^2 /Q^2 \ll
1$. Within them, the longitudinal structure function $F_{UU ,L}$ in the
numerator of Eq.~\eqref{e:mFF} can be neglected~\cite{Bacchetta:2008xw}. In
the denominator, the standard inclusive longitudinal structure function $F_L$
is non negligible and contains contributions of order $\alpha_S$. However, in
our analysis we assume a parton-model picture and we neglect such
contributions; hence, consistently we neglect the contribution of $F_L$  in
the denominator of Eq.~\eqref{e:mFF}.  
It may also be noted that in the transverse-momentum analysis of the data,
$F_{L}$ induces a change in normalization that depends on $x$, but is
independent of $z$ and 
$\bm{P}_{h\Tperp}^2$, the kinematic variables most relevant in the fitting
procedure. Hence, we do not expect large effects on the resulting
parameters. 

To express the structure functions in terms of TMD PDFs and FFs, 
we rely on the factorized formula 
for semi-inclusive DIS at low transverse  
momenta~\cite{Collins:1981uk,Collins:1984kg,Ji:2002aa,Ji:2004wu,%
Collins:2011zzd,Aybat:2011zv,GarciaEchevarria:2011rb,Echevarria:2012pw,%
Collins:2012uy}:  
\begin{align}
\label{e:parton2}
   F_{UU,T}(x,z, \bm{P}_{h \Tperp}^2, Q^2) &= \sum_a \mathcal{H}_{UU,T}^{a}(Q^2;\mu^2) \, 
      \int d\bm{k}_\T^{} \, d\bm{P}_\T^{} \,  f_1^a\big(x,\bm{k}_{\T}^2; \mu^2 \big) \, D_{1}^{a\smarrow h}\big(z,\bm{P}_{\T}^2; \mu^2 \big) \,
      \delta \big(z {\bm k}_{\T} - {\bm P}_{h \Tperp} + {\bm P}_{\T}\big)
\nonumber\\&
 + Y_{UU,T}\big(Q^2, \bm{P}_{h\Tperp}^2\big) + \mathcal{O}\big(M/Q\big) \, .
\end{align} 
Here, $\mathcal{H}_{UU,T}$ is the hard scattering part; $f_1^a(x,\bm{k}_{\T}^2;
\mu^2)$ is the TMD PDF for an unpolarized parton of flavor $a$ in an unpolarized
proton, carrying longitudinal momentum fraction $x$ and transverse momentum
$\bm{k}_\T$ at the factorization scale $\mu^2$, which in the following we
choose to be equal to $Q^2$.  $D_1^{a\smarrow h}(z, \bm{P}_{\T}^2;
\mu^2)$ is the TMD FF for an unpolarized parton of flavor $a$ fragmenting into
an unpolarized hadron $h$ carrying longitudinal momentum fraction $z$ and
transverse momentum 
$\bm{P}_\T$; the term $Y_{UU,T}$ is introduced to ensure a matching
to the perturbative
calculations at high transverse momentum. The
expression for $F_{UU,T}$ is known up to at least  ${\cal O}(\alpha_S^2)$,
including the resummation of at least next-to-next-to-leading logarithms of
the type $\log{(P_{h \Tperp}^2/Q^2)}$. However,  we are going to use here only
the lowest-order expression, which should still provide a good description at
low $\bm{P}_{h\Tperp}^2$ and in a limited range of $Q^2$. Eventually,
Eq.~\eqref{e:parton2} simplifies to (see, e.g.,
Refs.~\cite{Mulders:1996dh,D'Alesio:2004up,Bacchetta:2006tn}) 
\begin{equation} 
\label{e:F_UUT_simpl}
F_{UU ,T} (x,z,\bm{P}_{h \Tperp}^2, Q^2) =  \sum_a \, e_a^2 \; \bigl[ f_1^a \otimes D_1^{a\smarrow h} \bigr] (x,z,\bm{P}_{h \Tperp}^2, Q^2)  \, ,
\end{equation}
where the convolution upon transverse momenta is defined as 
\begin{equation}
\bigl[ f \otimes D \bigr] (x,z,\bm{P}_{h \Tperp}^2, Q^2)  = x \int d\bm{k}_\T^{}\,  d\bm{P}_\T^{}\, 
\delta \bigl(z \bm{k}_\T + \bm{P}_\T^{} - \bm{P}_{h \Tperp} \bigr)\, f(x,\bm{k}_\T^2; Q^2)\, D(z,\bm{P}_\T^2; Q^2) \, .
\label{e:convo}
\end{equation}

\begin{figure}
\centering
\includegraphics[width=10cm]{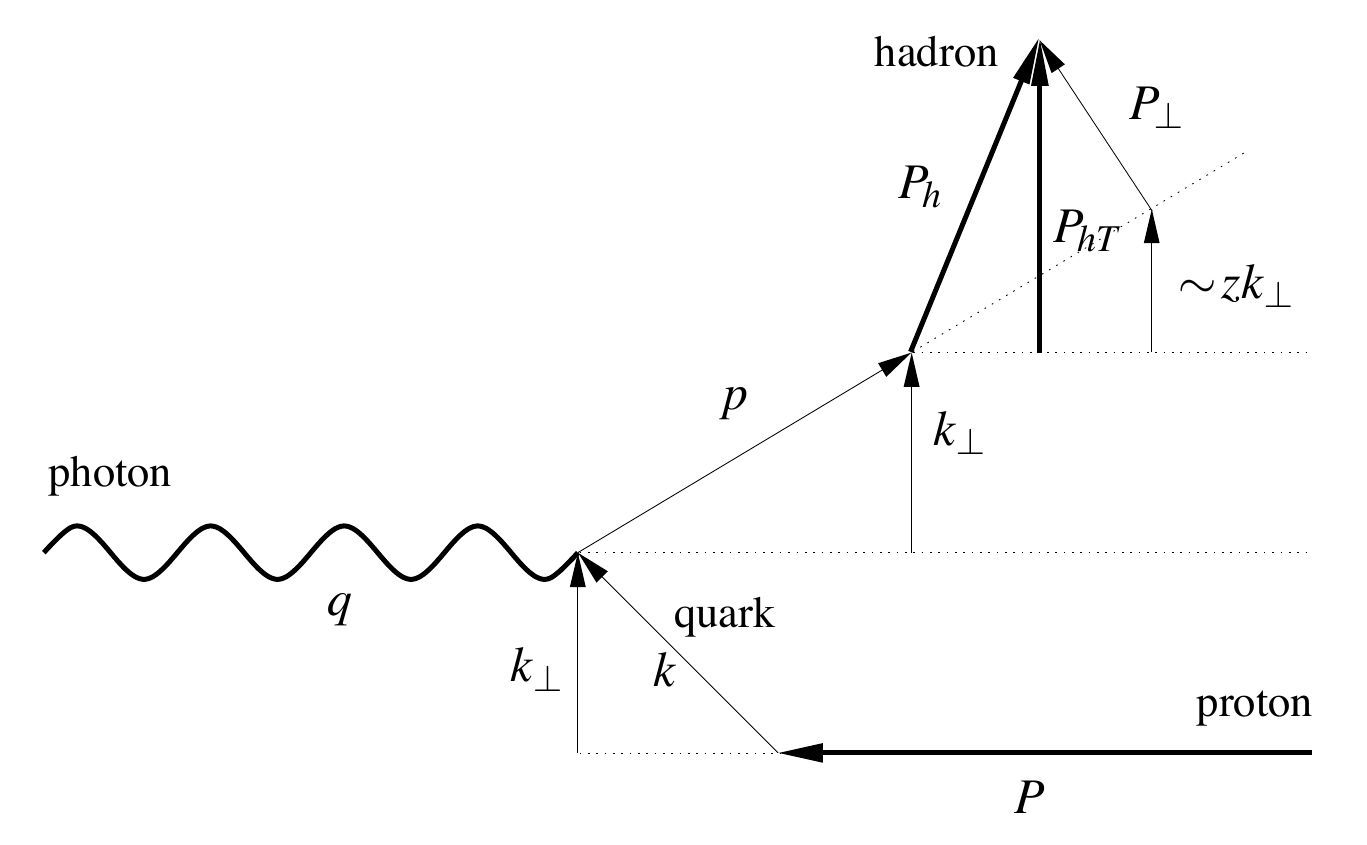}
\caption{Diagram describing the relevant momenta involved in a semi-inclusive DIS event: a
  virtual photon (defining the reference axis) strikes a parton inside a
  proton. The parton has a transverse momentum $\bm{k}_\T$ (not measured). The
  struck parton fragments into a hadron, which acquires a further transverse
  momentum $\bm{P}_\T$ (not measured). 
 The total measured transverse-momentum of the
  final hadron is $\bm{P}_{h\Tperp}$. When $Q^2$ is very large, the longitudinal
  components are all much larger than the transverse components. In this
  regime,  
  $\bm{P}_{h\Tperp} \approx z \bm{k}_\T + \bm{P}_\T$ (see also
  Ref.~\cite{Rajotte:2010}).} 
\label{f:trans_momenta}
\end{figure}

In Fig.~\ref{f:trans_momenta}, we describe our notation for the transverse momenta (in agreement with the notation suggested by the white paper in 
Ref.~\cite{Boer:2011fh}), which is also reproduced below for convenience:\\
\begin{center}
\begin{tabular}{ll}
Momentum  &  Physical description \\
\hline
$k$ & 4-momentum of parton in distribution function 
\\
$p$ & 4-momentum of fragmenting parton 
\\
$\bm{k}_\T$ & light-cone transverse momentum of parton in distribution function  
\\ 
 $\bm{P}_\T$ & light-cone transverse momentum of final hadron w.r.t. fragmenting parton
\\
$\bm{P}_{h\Tperp}$ & light-cone transverse momentum of final hadron w.r.t. virtual photon 
\end{tabular} 
\end{center}

%%%
%%%

%%%%%%%%%%%
\subsection{Flavor-dependent Gaussian ansatz}

The Gaussian ansatz consists in assuming the following functional form for the
transverse-momentum dependence of both 
the TMD PDF $f_1^a$ and the TMD FF $D_1^{a\smarrow h}$ in Eq.~\eqref{e:F_UUT_simpl}:
\begin{align}
f_1^a(x,\bm{k}_\T^2; Q^2) &= \frac{f_1^a(x,Q^2)}{\pi \langle \bm{k}_{\T,a}^2 \rangle}\; e^{-\bm{k}_\T^2/\langle \bm{k}_{\T,a}^2 \rangle} 
&
D_1^{a\smarrow h}(z,\bm{P}_\T^2, Q^2) &= \frac{D_1^{a\smarrow h}(z;Q^2)}{\pi \langle \bm{P}_{\T,a\smarrow h}^2 \rangle}\; 
e^{- \bm{P}_\T^2/\langle \bm{P}_{\T,a\smarrow h}^2 \rangle} \, .
\label{e:fldep_gauss}
\end{align} 
Due to its simplicity, this ansatz has been widely used in phenomenological studies but with constant widths 
$\langle \bm{k}_\T^2 \rangle$ and $\langle \bm{P}_\T^2 \rangle$. Here, for the first time we introduce an explicit dependence on flavor $a$ for both average transverse momenta $\langle \bm{k}_{\T,a}^2 \rangle$ and $\langle \bm{P}_{\T,a\smarrow h}^2 \rangle$. In principle, there are no reasons to prefer the Gaussian ansatz over other functional forms, and indeed more flexible forms should be investigated in the future. Model calculations typically lead to a non-Gaussian 
behavior~\cite{Jakob:1997wg,Bacchetta:2008af,Pasquini:2008ax,Wakamatsu:2009fn,Lorce:2011dv,Avakian:2010br}. 
The ansatz is also not compatible with the proper QCD evolution of TMD PDFs: it could be at most applicable at one specific starting scale, but would soon be spoiled by QCD corrections. In our analysis, we completely neglect $Q^2$ evolution, even in the collinear part of the functions, which we evaluate at $Q^2 = 2.4$ GeV$^2$. We can do this only because the range in $Q^2$ spanned by the \hermes measurements is not large. From now on, we drop the $Q^2$ dependence of the involved functions.

The convolution on transverse momenta in Eq.~\eqref{e:convo} can be solved analytically:
\begin{equation} 
\begin{split} 
\bigl[ f_1^a \otimes D_1^{a\smarrow h} \bigr] (x,z,\bm{P}_{h\Tperp}^2) &=   f_1^a(x) \, D_1^{a\smarrow h}(z)\, 
\bigg[ \frac{e^{- \bm{k}_\T^2/\langle \bm{k}_{\T,a}^2 \rangle}}{\pi \langle \bm{k}_{\T,a}^2 \rangle} \otimes 
            \frac{e^{- \bm{P}_\T^2/\langle \bm{P}_{\T,a\smarrow h}^2 \rangle}}{\pi \langle \bm{P}_{\T,a\smarrow h}^2 \rangle}  \bigg]
\\
&= x \, f_1^a(x) \, D_1^{a\smarrow h}(z)\, \frac{1}{\pi \langle \bm{P}_{h\Tperp,a}^2 \rangle}\; e^{-\bm{P}_{h\Tperp}^2/\langle \bm{P}_{h\Tperp,a}^2 \rangle} \, , 
\end{split}  
\label{e:convsolve}
\end{equation} 
where the relation between the three variances is
\begin{equation}
\langle \bm{P}_{h\Tperp,a}^2 \rangle = z^2 \langle \bm{k}_{\T,a}^2 \rangle + \langle \bm{P}_{\T,a\smarrow h}^2 \rangle \, .
\label{e:fldep_transvmom_rel}
\end{equation}
In this way, for each involved flavor $a$ the average square value of the transverse momentum $\bm{P}_{h\Tperp}$ of the detected hadron $h$ can be related to the average square values of the intrinsic transverse momenta $\bm{k}_\T$ and $\bm{P}_\T$, not directly accessible by experiments. 

Inserting Eq.~\eqref{e:convsolve} in Eq.~\eqref{e:F_UUT_simpl}, we simplify the multiplicities as
\begin{equation}
\begin{split}
m_{N}^h (x,z,\bm{P}_{h\Tperp}^2) &= \frac{ \pi }{ \sum_{a} e_a^2 \, f_{1}^{a} (x) }  \\
& \times \sum_{a} e_a^2 \; f_{1}^{a} (x) \, D_{1}^{a \smarrow h} (z)\; 
\frac{ e^{ - { \bm{P}_{h\Tperp}^2 }/\big( z^2 \langle \bm{k}_{\T,a}^2
      \rangle + \langle \bm{P}_{\T,a\smarrow h}^2 \rangle\big) } }
        { \pi \big(z^2 \langle \bm{k}_{\T,a}^2 \rangle + \langle \bm{P}_{\T,a\smarrow h}^2 \rangle\big) }  \,  .
\label{e:FDmult}
\end{split}  
\end{equation} 
If the distribution 
%and fragmentation 
functions describe a parton $a$ in a proton target, obviously the above
expression is valid for $N=p$, i.e.,  for a proton target. We can deduce the
corresponding result for a neutron target by assuming isospin symmetry. For a
deuteron target, we can assume an incoherent sum of proton and neutron
contributions.  
Under these assumptions the necessary label for the parent hadron
on PDFs is omitted and  
PDFs refer to the ones of the proton.  
We remark also that each quark flavor is described by a single Gaussian with a
specific width. The multiplicity is then a sum of Gaussians and thus no
longer a simple Gaussian. The above expression can be used with minor
modifications also if we assume that the distribution and fragmentation
functions for some flavor are themselves sums of Gaussians. We will in fact
adopt such an assumption for the up and down quarks, where we distinguish a
valence and a sea contribution, each one having a different Gaussian
width. For example, the up contribution to the multiplicities is 
\begin{equation} 
\begin{split} 
\bigl[ f_1^u\otimes D_1^{u\smarrow h} \bigr] &(x,z,\bm{P}_{h\Tperp}^2) = 
\bigl[ (f_1^{u_v} + f_1^{\bar{u}}) \otimes D_1^{u\smarrow h} \bigr] (x,z,\bm{P}_{h\Tperp}^2)  
\\ 
& = x \, f_{1}^{u_v} (x) \, D_{1}^{u \smarrow h} (z)\; 
           \frac{ e^{ - { \bm{P}_{h\Tperp}^2 }/\big( z^2 \langle \bm{k}_{\T,u_v}^2 \rangle + \langle \bm{P}_{\T,u\smarrow h}^2 \rangle \big)} }
                   { \pi \big(z^2 \langle \bm{k}_{\T,u_v}^2 \rangle + \langle
                     \bm{P}_{\T,u\smarrow h}^2 \rangle \big) }     +
x \, f_{1}^{\bar{u}} (x) \, D_{1}^{u \smarrow h} (z)\; 
\frac{ e^{ - { \bm{P}_{h\Tperp}^2 }/\big( z^2 \langle \bm{k}_{\T,\bar{u}}^2 \rangle + \langle \bm{P}_{\T,u\smarrow h}^2 \rangle \big)} }
        { \pi \big(z^2 \langle \bm{k}_{\T,\bar{u}}^2 \rangle + \langle
          \bm{P}_{\T,u\smarrow h}^2 \rangle\big) } \, , 
\end{split}  
\end{equation} 
where $f_{1}^{u_v} = f_{1}^{u}- f_{1}^{\bar{u}}$, and similarly for the down quark.

Previous data obtained in unpolarized Drell-Yan and semi-inclusive DIS processes were
compatible with calculations based on a Gaussian ansatz for unpolarized TMD
PDFs and TMD FFs with flavor-independent constant widths. In this case,
Eq.~\eqref{e:FDmult} would display a simple Gaussian behavior in $\bm{P}_{h\Tperp}$
with the same width in every target-hadron combination.  
However, the \hermes multiplicities display significant differences between
proton and deuteron targets, and between pion and kaon final-state
hadrons. Hence, they strongly motivate our choice in Eq.~\eqref{e:fldep_gauss}
for a flavor-dependent Gaussian ansatz.

%%%%%%%%%%%
\subsection{Assumptions concerning average transverse momenta}

As mentioned in the previous section, we introduce different widths for the
Gaussian forms of the valence and sea components of up and down TMD
PDFs. However, we assume that the Gaussian widths 
of all sea quarks ($\bar{u}$, $\bar{d}$, $s$ and $\bar{s}$) are the same (i.e., they have the same average
square transverse momenta).
State-of-the-art parametrizations of
collinear PDFs have a more complex structure and introduce differences between
sea quarks of different flavors; we leave this flexibility to future studies. 

We include the possibility that the average square transverse momentum depends
on the longitudinal fractional momentum $x$. This connection can certainly be
useful in fitting the data, but above all it is dictated by theoretical
considerations, in particular by Lorentz invariance. Many models 
predict such a connection (see,  
e.g.,~\cite{Bacchetta:2008af,Bacchetta:2010si,Wakamatsu:2009fn,Efremov:2010mt,%
Bourrely:2010ng,Matevosyan:2011vj,Schweitzer:2012hh,%
Pasquini:2008ax,Lorce:2011dv,Avakian:2010br}),
and similarly do parametrizations of light-front wave functions (see,
e.g.,~\cite{Brodsky:2000ii,Hwang:2007tb,Gutsche:2013zia}).  

We choose the following functional form for the average square transverse momentum of flavor $a$:
\begin{align} 
\big\langle \bm{k}_{\T,a}^2 \big\rangle (x) = 
\big\langle \hat{\bm{k}}_{\T,a}^2 \big\rangle \;  
\frac{(1-x)^{\alpha} x^{\sigma} }{ (1-\hat{x})^{\alpha} \hat{x}^{\sigma} } \, ,
\label{e:kT2_kin}
&&
\text{where }
\big\langle \hat{\bm{k}}_{\T,a}^2 \big\rangle\equiv \big \langle \bm{k}_{\T,a}^2
\big \rangle
(\hat{x}),
\text{ and }
\hat{x}=0.1.
\end{align} 
$\langle \hat{\bm{k}}_{\T,a}^2 \rangle$, $\alpha$, $\sigma$, are free
parameters.
For sake of simplicity, we keep the same exponents $\alpha$ and $\sigma$ for
all flavors. According to the above assumptions, we have three more
parameters:  
$\langle \hat{\bm{k}}_{\T,a}^2 \rangle$ for $a = u_v,\, d_v,\, {\rm sea}$. 
In total, we use five different parameters to describe all TMD PDFs.
Since the present data have a limited coverage in $x$, we found no need of
more sophisticated choices.  

As for  TMD FFs, fragmentation processes in which the fragmenting parton is in
the valence content of the detected hadron are usually defined {\em
  favored}. Otherwise the process is classified as {\em unfavored}. The
biggest difference between the 
two classes is the number of $q\bar{q}$ pairs excited from the vacuum in order
to produce the detected hadron: favored processes involve the creation of at
most one $q\bar{q}$ pair. If the final hadron is a kaon, we further
distinguish a favored process initiated by a strange quark/antiquark from a
favored process initiated by an up quark/antiquark. 

For simplicity, we assume charge conjugation and isospin symmetries. The
latter is often imposed also in the parametrization of 
collinear FFs~\cite{Hirai:2007cx}, but not always~\cite{deFlorian:2007aj}. In
practice, we consider four different Gaussian shapes: 
\begin{gather}
\big\langle \bm{P}^2_{\T,u \smarrow \pi^+} \big\rangle = \big \langle
\bm{P}^2_{\T,\bar{d} \smarrow \pi^+} \big \rangle = \big \langle \bm{P}^2_{\T,\bar{u} \smarrow \pi^-} \big \rangle =  \big \langle \bm{P}^2_{\T,d \smarrow \pi^-}\big \rangle \equiv \big \langle \bm{P}^2_{\T, {\rm fav}} \big \rangle \, ,  
\label{e:favored}  
\\
\big \langle \bm{P}^2_{\T,u \smarrow K^+} \big \rangle =  \big \langle \bm{P}^2_{\T,\bar{u} \smarrow K^-} \big \rangle \equiv \big \langle \bm{P}^2_{\T, {uK}} \big \rangle \, ,
\label{e:uK}  
\\
\big \langle \bm{P}^2_{\T,\bar{s} \smarrow K^+} \big \rangle = \big \langle \bm{P}^2_{\T,s \smarrow K^-}\big \rangle \equiv \big \langle \bm{P}^2_{\T, {sK}} \big \rangle \,  ,
\label{e:sK}  
\\
\big \langle \bm{P}^2_{\T,\text{all others}} \big \rangle 
%\big \langle P_{\T,\bar{u} \smarrow \pi^+} \big \rangle =
%\big \langle P_{\T,\bar{d} \smarrow \pi^-} \big \rangle = \big \langle P_{\T, u \smarrow \pi^-} \big \rangle = 
%\big \langle P_{\T, (s,b,c) \smarrow \pi^+} \big \rangle = \big \langle P_{\T, (s,c,b) \smarrow \pi^-} 
\equiv  \big \langle \bm{P}^2_{\T, {\rm unf}} \big \rangle \, .
\label{e:unfavored} 
\end{gather} 
The last assumption is made mainly to keep the number of parameters under
control, though it could be argued that unfavored fragmentation into kaons is
different from unfavored fragmentation into pions.

As for TMD PDFs, also for TMD FFs we introduce a dependence of the average
square transverse momentum on the longitudinal momentum fraction $z$, as done
in several models or phenomenological
extractions (see, e.g., Refs.~\cite{Boglione:1999pz,
Schweitzer:2003yr,D'Alesio:2004up,Bacchetta:2002tk,Bacchetta:2007wc,%
Matevosyan:2011vj}).  We
choose the functional form 
\begin{align}  
\big \langle \bm{P}_{\T,a \smarrow h}^2 \big \rangle (z) &= \big \langle
\hat{\bm{P}}_{\T,a \smarrow h}^2 \big \rangle 
\frac{ (z^{\beta} + \delta)\ (1-z)^{\gamma} }{ (\hat{z}^{\beta} + \delta)\
  (1-\hat{z})^{\gamma} } \, 
&&
\text{where }
\big\langle \hat{\bm{P}}_{\T,a \smarrow h}^2 \big\rangle\equiv \big \langle
\bm{P}_{\T,a \smarrow h}^2
\big \rangle
(\hat{z}),
\text{ and }
\hat{z}=0.5.
\label{e:PT2_kin}
\end{align}  
The free parameters $\beta$, $\gamma$, and $\delta$ are equal for all kinds of fragmentation
functions. In conclusion, we use seven different parameters to describe all
the TMD FFs.

%%%%%%%%%%%%%%%%%%%%%%%%%%%%%%%%%%%%%%%%%%%%
\section{Analysis procedure}
%%%%%%%%%%%%%%%%%%%%%%%%%%%%%%%%%%%%%%%%%%%%
\label{s:analysis}

\subsection{Selection of data}
\label{s:selection}

The \hermes collaboration collected a total of 2688 data points (336 points
for each of the 8 combination of target and final-state hadrons), with the
average values of $(x,Q^2)$ ranging from about $(0.04,1.25 \text{ GeV}^2)$ to
about $(0.4,9.2 \text{ GeV}^2)$, $0.1\leq z \leq 0.9$, and $0.1 \text{ GeV}\leq
|\bm{P}_{h \Tperp}| \leq 1$ GeV. The collaboration presented two distinct data
sets, including or neglecting vector meson contributions. Here, we use the
data set where the vector meson contributions have been subtracted.
In all
cases, we sum in quadrature statistical and systematic errors and we ignore
correlations. We always use the average values of the kinematic variables in
each bin.

Our analysis relies on the assumption that the transverse-momentum-integrated
multiplicities, $m_N^h(x,z,Q^2)$, 
are well described by currently available parametrizations of
collinear PDFs and FFs. However, this is not always true. In order to identify
the range of applicability of the collinear results, we compared the
multiplicities as functions of $x$ and $z$ with the leading-order (LO)
theoretical predictions obtained using the MSTW08LO PDF
set~\cite{Martin:2009iq} and the DSS LO FF set~\cite{deFlorian:2007aj}.  
In the comparison, we neglected the uncertainties affecting the PDFs but we
included the ones affecting the FFs, obtaining the latter from the plots in
Ref.~\cite{Epele:2012vg}. They are of the order of 5-10\% for light quarks
fragmenting into pions, of 10-15\% for favored kaon FFs, of 50\% for all the
other cases, and they are larger at higher $z$.  

\renewcommand{\arraystretch}{2}

%%%%%%%%%%%%%%%   Tab chi2 collinear   %%%%%%%%%%%%%%%
\begin{table}
  \centering
  \begin{tabular}{|c||c|c|c|c|c|c|c|c|c|}
  \hline
  \multicolumn{10}{|c|}{$\chi^2$/d.o.f.} \\
  \hline
  \hline 
                   &  global   &     $p \to K^-$   &  $p \to \pi^-$  &  $p \to \pi^+$  &  $p \to K^+$  &  $D \to K^-$   &  $D \to \pi^-$  &  $D \to \pi^+$  &  $D \to K^+$  \\
  \hline
$Q^2 >1.4$ GeV$^2$  
   & 2.86 & 2.25 & 3.39  & 1.87 & 0.89 & 4.26 & 5.05 & 3.33 & 1.80 
\\
\parbox{25mm}{$Q^2 >1.4$ GeV$^2$\\  (no VM subtr.)}
   & 3.90 & 2.27 & 6.58  & 2.45 & 0.85 & 4.22 & 8.66 & 4.61 & 1.57 
\\
\parbox{25mm}{$Q^2 >1.4$ GeV$^2$\\  (with evolution)}
   & 3.55 & 1.38 & 5.03  & 2.74 & 1.13 & 2.81 & 7.96 & 5.19 & 2.17 
\\
$Q^2 >1.6$ GeV$^2$
   & 2.29 & 2.38 & 2.70  & 1.16 & 0.59 & 4.45 & 3.42 & 2.29 & 1.31 
\\
\hline
\end{tabular}
\caption{Values of $\chi^2/{\rm d.o.f.}$ obtained from the comparison of the \hermes multiplicities $m_N^h(x,z,Q^2)$ with the theoretical prediction using the MSTW08LO collinear PDFs~\cite{Martin:2009iq} and the DSS LO collinear 
FFs~\cite{deFlorian:2007aj}. In all cases, the range $0.1 \leq z \leq 0.8$ was included.}
\label{t:chi2collinear}
\end{table}
 
In Tab.~\ref{t:chi2collinear}, we quote the $\chi^2$ per degree of freedom ($\chi^2/\text{d.o.f.}$) obtained in our comparison. Our results are different from the ones quoted in Tabs.~IV and VIII of~\cite{deFlorian:2007aj} for a few reasons: i) we used the final \hermes data, in particular the set with $x$ and $z$ binning; ii) we included also the lowest $z$ bin ($z<0.2$);  iii) we did not include any overall normalization constant; iv) we included the theoretical errors on the extracted fragmentation functions. 
The comparison shows that in general the theoretical predictions do not
describe the \hermes data very well. The agreement is particularly bad
for $\pi^-$ and $K^-$. However, this is not surprising because: i) the MSTW
set of PDFs does not take into account semi-inclusive DIS data, ii) as mentioned above, the
DSS set of FFs~\cite{deFlorian:2007aj} was deduced using only a preliminary
version of the \hermes multiplicities, iii) the \hermes data give very large
contributions to the $\chi^2$ of the global DSS analysis.  
Nevertheless, in our analysis we decided to restrict the ranges to $Q^2>1.4$
GeV$^2$ and $0.1 < z < 0.8 $, i.e.,  excluding the first bin in $Q^2$
(equivalent also to the lowest $x$) and the last bin in $z$.
 Inclusion
of decays from exclusive vector-mesons markedly degrades the $\chi^2$ of the
pion channels and increases the global $\chi^2$
(cf.\ the first and second line of Tab.~\ref{t:chi2collinear}). 
Hence we will present results for only
the fits to vector-meson subtracted multiplicities. We checked that
our basic conclusions do not change when using data without vector-meson subtraction. 

We also noted that a description of data of comparable quality could be
achieved by turning off the $Q^2$ dependence of both collinear PDFs and FFs,
and by computing them at the fixed value of $Q^2=2.4$ GeV$^2$ (cf.\ the first
and third line of Tab.~\ref{t:chi2collinear}). 
Therefore, we
decided to systematically neglect any contribution of QCD evolution and to
compute all theoretical quantities at the average value of $Q^2=2.4$ GeV$^2$. 

When considering also the transverse-momentum dependence, the TMD formalism is valid only when 
$\bm{P}_{h \Tperp}^2 \ll Q^2$. In order not to exclude too many data points, we apply the loose requirement 
$\bm{P}_{h \Tperp}^2 < Q^2/3$. This leads to the exclusion of at most two
bins at high $\bm{P}_{h \Tperp}^2$ and low $Q^2$.  

Finally, we exclude also the data points with the lowest $|\bm{P}_{h\Tperp}|$
($|\bm{P}_{h\Tperp}| <0.15$ GeV).  
A priori, there is no reason to exclude them, but in our attempts
we found them particularly difficult to fit, mainly because they often do not
follow the trend of the other data points and at the same time they have small
errors. In order to be able to fit them, we need to increase the flexibility
of our functional forms. We leave this task to future studies. 

In summary, we use a total of 1538 data points, approximately 190 for each of
the 8 combinations of target and final-state hadrons, which correspond to 
about 60\% of the
total 2688 points measured by the \hermes collaboration.

%%%%%%%%%%%
\subsection{Fitting procedure and uncertainties}

The fit and the error analysis were carried out using a similar Monte Carlo approach as in Ref.~\cite{Bacchetta:2012ty}, and taking
inspiration from the work of the NNPDF collaboration (see, e.g.,
\cite{Forte:2002fg,Ball:2008by,Ball:2010de}). The approach consists in
creating $\mathcal{M}$ replicas of the data points. In each replica (denoted
by the index $r$), each data point $i$ is shifted by a Gaussian noise with the
same variance as the measurement. Each replica, therefore, represents a
possible outcome of an independent experimental measurement, which we denote
by $m_{N, r}^{h}(x, z, \bm{P}_{h\Tperp}^2, Q^2)$. The number of replicas is
chosen so that the mean and standard deviation of the set of replicas
accurately reproduces the original data points. In our case, we have found
that 200 replicas are more than sufficient.

The standard minimization procedure is applied to each replica separately, by
minimizing the following error  
function~\footnote{Note that the error for each replica is taken to be equal
  to the error on the original data points. This is consistent with the fact
  that the variance of the $\mathcal{M}$ replicas should reproduce the
  variance of the original data points.}  
\begin{equation}
E_r^2(\{p\})=\sum_{i} 
\frac{\Bigl(m_{N, r}^{h}(x_i, z_i, \bm{P}_{h\Tperp i}^2, Q_i^2) - m_{N,  \mbox{\tiny theo}}^{h}(x_i, z_i, \bm{P}_{h\Tperp i}^2; \{p\})\Bigr)^2}
        {\Bigl(\Delta m_{N, \mbox{\tiny stat}}^{h}(x_i, z_i, \bm{P}_{h\Tperp i}^2, Q^2_i) \Bigr)^2+\Bigl(\Delta m_{N, \mbox{\tiny sys}}^{h}(x_i, z_i, \bm{P}_{h\Tperp i}^2, Q^2_i) \Bigr)^2+\Bigl(\Delta m_{N, \mbox{\tiny theo}}^{h}(x_i, z_i, \bm{P}_{h\Tperp i}^2) \Bigr)^2}  \, . 
\label{e:MC_chi2}
\end{equation}
The sum runs over the $i$ experimental points, including all species of
targets $N$ and final-state hadrons $h$. The theoretical multiplicities $m_{N,
  \text{theo}}^{h}$ and their error $\Delta m_{N,  {\rm theo}}^{h}$ do not
depend on $Q^2$, as explained in the previous section. They are computed at
the fixed value $Q^2=2.4$ GeV$^2$ using the formula in
Eq.~\eqref{e:FDmult}. However, in each $z$ bin for each replica the value of
$D_1^{a \smarrow h}$ is independently modified with a Gaussian noise with
standard deviation equal to the theoretical error $\Delta D_1^{a\smarrow
  h}$. The latter is estimated from the plots in Ref.~\cite{Epele:2012vg} and
it represents the main source of uncertainty in $\Delta m_{N,  {\rm
    theo}}^{h}$. Finally, the symbol $\{p\}$ denotes the vector of fitting
parameters.  

The minimization was carried out using the \minuit  code. The final outcome is a set of $\mathcal{M}$ different vectors of best-fit parameters, $\{ p_{0r}\},\; r=1,\ldots \mathcal{M}$, with which we can calculate any observable, its mean, and its standard deviation. 
The distribution of these values needs not to be necessarily Gaussian. In this
case, the $1 \sigma$ confidence interval is different from the 68\%
interval. The 68\% confidence interval can simply be computed for
each experimental point by rejecting the largest and the lowest 16\% of the
$\mathcal{M}$ values.   

Although the minimization is performed on the function defined in Eq.~\eqref{e:MC_chi2}, the agreement of the $\mathcal{M}$ replicas with the original data is better expressed in terms of a $\chi^2$ function defined as in Eq.~\eqref{e:MC_chi2} but with the
replacement $m_{N, r}^{h} \to m_{N}^{h}$, i.e.,  with respect to the original data set. If the model is able to give a good
description of the data, the distribution of the $\mathcal{M}$ values of
$\chi^2$/d.o.f. 
should be peaked around one.  In practice, the rigidity of our
functional form leads to higher $\chi^2$ values.

%%%%%%%%%%%%%%%%%%%%%%%%%%%%%%%%%%%%%%%%%%%%
\section{Results}
%%%%%%%%%%%%%%%%%%%%%%%%%%%%%%%%%%%%%%%%%%%%
\label{s:results}

In this section, we describe the results obtained by fitting the \hermes multiplicities with the theoretical formula of 
Eq.~\eqref{e:FDmult} and using the Monte Carlo method outlined in the previous section. We performed different kinds
of fits with different assumptions. The first one, conventionally named
``default fit," includes all the 1538 data points selected according to the
criteria described in Sec.~\ref{s:selection}. In the second one, we excluded 
data also for the second lowest $Q^2$ bin, i.e., by selecting $Q^2
> 1.6$ GeV$^2$. This cut reduces the number of data points to 1274. The third
scenario corresponds to neglecting kaons in the final state and taking only
multiplicities for pions. The last scenario is a fit of the default selection
using a flavor-independent Gaussian ansatz. Before discussing each
different scenario, here below we list their common features. 

As repeatedly mentioned above, in our analysis we neglected completely the
effect of $Q^2$ evolution,  even in the collinear 
PDFs and FFs, and we evaluated all observables at the fixed value $Q^2 = 2.4$
GeV$^2$.  
 
As for the dependence of the TMD average transverse momentum on $x$, we
noticed that the fit is weakly sensitive to the exponents in
Eq.~\eqref{e:kT2_kin}. We tried fits with $\alpha= \sigma=0$ and obtained good
results. However, in order to stress the fact that present data do not
constrain these parameters very well, we decided to assign random values
extracted from uniform distributions to both the exponents: we consider
$\alpha$ as a random number between 0 and 2 and $\sigma$ as a random number
between $-0.3$ and $0.1$ . Better determinations of these parameters require
an extended range in $x$, together with uncorrelated $x$ and $Q^2$
binnings. The dependence of the TMD FF average transverse momentum on $z$ is
governed by Eq.~\eqref{e:PT2_kin}; in this case, we decided to keep all three
parameters free.  

For each scenario, we performed 200 replicas of the fit. In this section, we
present the 68\% confidence intervals of the parameters
over the replicas, computed by rejecting the largest and the lowest 16\% of the
replicated parameter values. We quote the values as $A \pm B$, where $A$ is
the average of the upper and lower limits of the 68\% confidence
interval and $B$ is their semi-difference.
It is understood that much more information is available by
scrutinizing the full set of 200 values for each of them.\footnote{The results
  will be available via the website http://tmd.hepforge.org or upon request.}  
In Tab.~\ref{t:fd_chi20dof}, we list the 68\% confidence intervals 
of the $\chi^2/$d.o.f.  for the different
scenarios, including the global result and the outcome for each target-hadron
combination, separately. In Tabs.~\ref{t:fd_PDFs_par} and \ref{t:fd_FFs_par},
we list the 68\% confidence intervals for the five fitting
parameters for TMD PDFs and for the seven fitting parameters for TMD FFs,
respectively.

%%%%%%%%%%%%%%%   Tab FD chi20dof   %%%%%%%%%%%%%%%
\begin{table}[t]
  \centering
  \begin{tabular}{|l||c|c|c|c|c|c|c|c|c|}
  \hline
  \multicolumn{10}{|c|}{$\chi^2$/d.o.f.} \\
  \hline
  \hline 
                   &  global   &     $p \to K^-$   &  $p \to \pi^-$  &  $p \to \pi^+$  &  $p \to K^+$  &  $D \to K^-$   &  $D \to \pi^-$  &  $D \to \pi^+$  &  $D \to K^+$  \\
  \hline
  Default           &$1.63\pm 0.12$ & $0.78\pm 0.15$ & $1.80\pm 0.27$ &
  $2.64\pm 0.21$ & 
$0.46\pm 0.07$ & $2.77\pm 0.56$ & $1.65\pm 0.20$ & $2.16\pm 0.21$ & 
$0.71\pm 0.15$ 
\\
  $Q^2 >1.6$ GeV$^2$&    $1.37\pm 0.12$ & $0.77\pm 0.14$ & $1.50\pm 0.24$ & $1.91\pm 0.30$ & 
$0.49\pm 0.07$ & $2.78\pm 0.52$ & $1.28\pm 0.19$ & $1.64\pm 0.25$ & 
$0.58\pm 0.12$ \\
  Pions only        &  $2.04\pm 0.16$ & --- & $1.68\pm 0.24$ & $2.70\pm 0.22$
  & --- & --- & $1.50\pm 0.18$ & $2.22\pm 0.22$ & --- \\
  Flavor-indep.&    $1.72\pm 0.11$ & $0.87\pm 0.16$ & $1.83\pm 0.25$ & $2.89\pm 0.23$ & 
$0.43\pm 0.07$ & $3.15\pm 0.62$ & $1.66\pm 0.20$ & $2.21\pm 0.22$ & 
$0.71\pm 0.15$ \\
\hline
\end{tabular}
\caption{68\% confidence intervals of $\chi^2/$d.o.f. values
  (global result and for every available target-hadron combination $N \to h$)
  for each of the considered four scenarios.} 
\label{t:fd_chi20dof}
\end{table}
%%%%%%%%%%%%%%%%%%%%%%%%%%%%%%%%%%%%%%%

%%%%%%%%%%%%   Tab PDFs parameters  %%%%%%%%%%%%%%
\begin{table}
  \centering
  \begin{tabular}{|c||c|c|c|c|c|}
  \hline
  \multicolumn{6}{|c|}{Parameters for TMD PDFs} \\
  \hline
  \hline 
                &  $\big \langle \hat{\bm{k}}_{\T, d_v}^2 \big \rangle$  &
                $\big \langle \hat{\bm{k}}_{\T, u_v}^2 \big \rangle$  &
                $\big \langle \hat{\bm{k}}_{\T, {\rm sea}}^2 \big \rangle$  &  $\alpha$  &
                $\sigma$      \\[-0.2cm]
                & [GeV$^2$] & [GeV$^2$] & [GeV$^2$] & (random) & (random)
\\
  \hline
  Default      &$0.30\pm 0.17$ & $0.36\pm 0.14$ & $0.41\pm 0.16$ & $0.95\pm 0.72$ & 
$-0.10\pm 0.13$     \\
  $Q^2 > 1.6\ \text{GeV}^2$	 &    
$0.33\pm 0.19$ & $0.37\pm 0.17$ & $0.31\pm 0.18$ & $0.93\pm 0.70$ & 
$-0.10\pm 0.13$ \\
  Pions only      	 &   $0.34\pm 0.12$ & $0.35\pm 0.12$ & $0.29\pm 0.13$
  & $0.95\pm 0.68$ &  $-0.09\pm 0.14$ \\
  Flavor-indep.	&    $0.30\pm 0.10$ & $0.30\pm 0.10$ & $0.30\pm 0.10$ &
 $1.03\pm 0.64$ & $-0.12\pm 0.12$ \\
\hline
\end{tabular}
\caption{68\% confidence intervals of 
best-fit parameters for TMD PDFs in the different scenarios.}
\label{t:fd_PDFs_par}
\end{table}
%%%%%%%%%%%%%%%%%%%%%%%%%%%%%%%%%%%%%%%

%%%%%%%%%%%%   Tab FFs parameters  %%%%%%%%%%%%%%
\begin{table}[h]
  \centering
  \begin{tabular}{|c||c|c|c|c|c|c|c|}
  \hline
  \multicolumn{8}{|c|}{Parameters for TMD FFs} \\
  \hline
  \hline 
     &  $\big \langle \hat{\bm{P}}_{\T, \text{fav}}^2 \big \rangle $  
          &  $\big \langle \hat{\bm{P}}_{\T, \text{unf}}^2 \big \rangle $  
          &  $\big \langle \hat{\bm{P}}_{\T, s K}^2 \big \rangle $  
          &  $\big \langle \hat{\bm{P}}_{\T, u K}^2 \big \rangle $  
&  $\beta$  &  $\delta$  &  $\gamma$      \\[-0.2cm]
                & [GeV$^2$] & [GeV$^2$] & [GeV$^2$] (random) &  [GeV$^2$] & &
                & \\
  \hline
  Default       &  $0.15\pm 0.04$ & $0.19\pm 0.04$ & $0.19\pm 0.04$ & $0.18\pm0.05$ 
  & $1.43\pm 0.43$ & $1.29\pm 0.95$ & $0.17\pm 0.09$ \\
  $Q^2 > 1.6\ \text{GeV}^2$	 &  $0.15\pm 0.04$ & $0.19\pm 0.05$ & $0.19\pm 0.04$ & $0.18\pm 0.05$ & $1.59\pm 0.45$ & $1.41\pm 1.06$ & $0.16\pm 0.10$ \\
  Pions only          		 &  $0.16\pm 0.03$ & $0.19\pm 0.04$ 
  & --- & --- &  $1.55\pm 0.27$ & $1.20\pm 0.63$ & $0.15\pm 0.05$ \\  
  Flavor-indep.	  &  $0.18\pm 0.03$ & $0.18\pm 0.03$ & $0.18\pm 0.03$ & $0.18\pm 0.03$ &
 $1.30\pm 0.30$ & $0.76\pm 0.40$ & $0.22\pm 0.06$ \\
\hline
\end{tabular}
\caption{68\% confidence intervals 
of best-fit parameters for TMD FFs in the different scenarios.}
\label{t:fd_FFs_par}
\end{table}
%%%%%%%%%%%%%%%%%%%%%%%%%%%%%%%%%%%%%%%

In all fits, we observe a strong anticorrelation between the distribution and
fragmentation transverse momenta. This is not surprising, since the width of
the observed $\bm{P}_{h\Tperp}$ distribution is given by
Eq.~\eqref{e:fldep_transvmom_rel}. To better pin down the values of $\langle
\bm{k}_{\T,a}^2 \rangle$ and  $\langle \bm{P}_{\T,a\smarrow h}^2 \rangle$
separately for the various flavors $a$, it will be essential to include also
data from electron-positron annihilations and Drell--Yan processes. In any
case, a common feature of all scenarios is that the 
$\langle \hat{\bm{k}}_{\T,  a}^2 \rangle$ (namely, the average squared
transverse momenta of TMD PDFs at 
$x=0.1$)  have average values around 0.3 GeV$^2$, while the $\langle
\hat{\bm{P}}_{\T, a\smarrow h}^2 \rangle$ (namely the average square
transverse momenta of TMD FFs at $z=0.5$) have average values around 0.18
GeV$^2$. Moreover, the fits prefer large values of the exponents $\beta$ and
$\delta$ for TMD FFs, but with large uncertainties; the parameter $\gamma$ is
usually small.

Here below, we discuss in detail the results for the four different scenarios.

\subsection{Default fit}

\begin{figure}
\centering
\includegraphics[width=7cm]{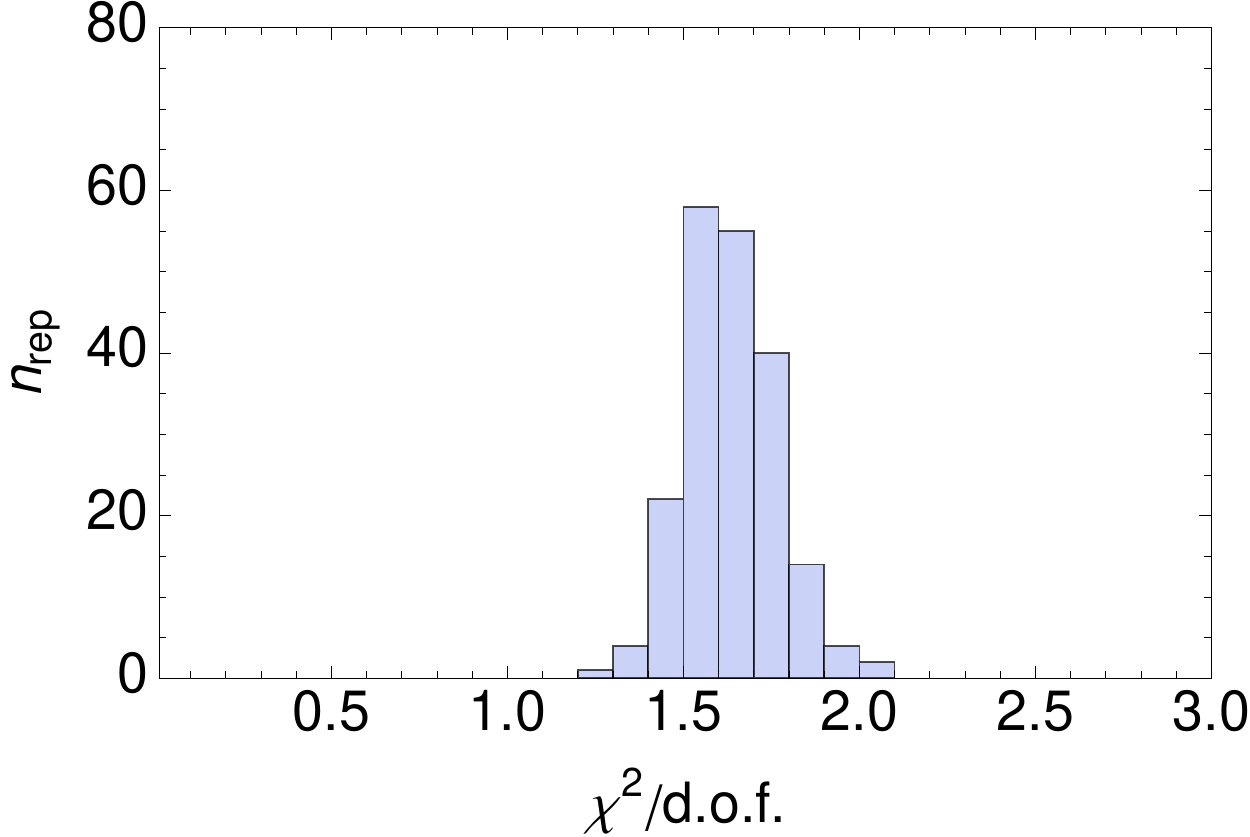}
\caption{Distribution of the values of $\chi^2/{\rm  d.o.f.}$ for the default fit. On the vertical axis, the number of replicas with 
$\chi^2/{\rm  d.o.f.}$ inside the bin. The bin width is $0.1$.}
\label{f:chi2distr}
\end{figure}

In this scenario, we consider all 1538 data points selected according to the criteria explained in Sec.~\ref{s:selection}. The quality of the fit is fairly good. The global $\chi^2/{\rm  d.o.f.}$ is $1.63 \pm 0.13$.  In Fig.~\ref{f:chi2distr}, the distribution of the 
$\chi^2/{\rm  d.o.f.}$ over the 200 replicas is shown. Many replicas have
$\chi^2/{\rm  d.o.f.} > 1.5$. This indicates some difficulty to reproduce the
data correctly. It is not surprising if we take into account that the
description of the collinear multiplicities was already difficult (see
Tab.~\ref{t:chi2collinear}). It may actually seem contradicting that our fit
is able to describe the transverse-momentum-dependent multiplicities
relatively well. This is probably simply due to the fact that the
multidimensional binning has many more data points but with much larger
statistical errors. 

%%%%

In Tab.~\ref{t:fd_chi20dof}, we list the 68\% confidence intervals 
of the $\chi^2/{\rm d.o.f.}$ also for each 
target-hadron combination $N \to h$, separately. The worst result is for $D \to K^-$. This may be a bit surprising, also because 
$p \to K^-$ is described very well. However, this may be due to the fact that
the collinear description of this channel is poor (see
Tab.~\ref{t:chi2collinear}). We point out also that the systematic errors in $D
\to K^-$ are significantly smaller than  
$p \to K^-$~\cite{Airapetian:2012ki}. The second worst agreement is for $p \to
\pi^+$, which is not unexpected since statistical errors are smallest in this
channel. The $\pi^-$ channels are described decently, which is at odds with
the poor description of their collinear multiplicities (see
Tab.~\ref{t:chi2collinear}). We do not have a reasonable explanation for this
feature yet. Maybe, it could be ascribed to the cuts in $\bm{P}_{h\Tperp}$ that
we implemented in our fit.  

%%%%%
\begin{figure}
\centering
\includegraphics[width=13cm]{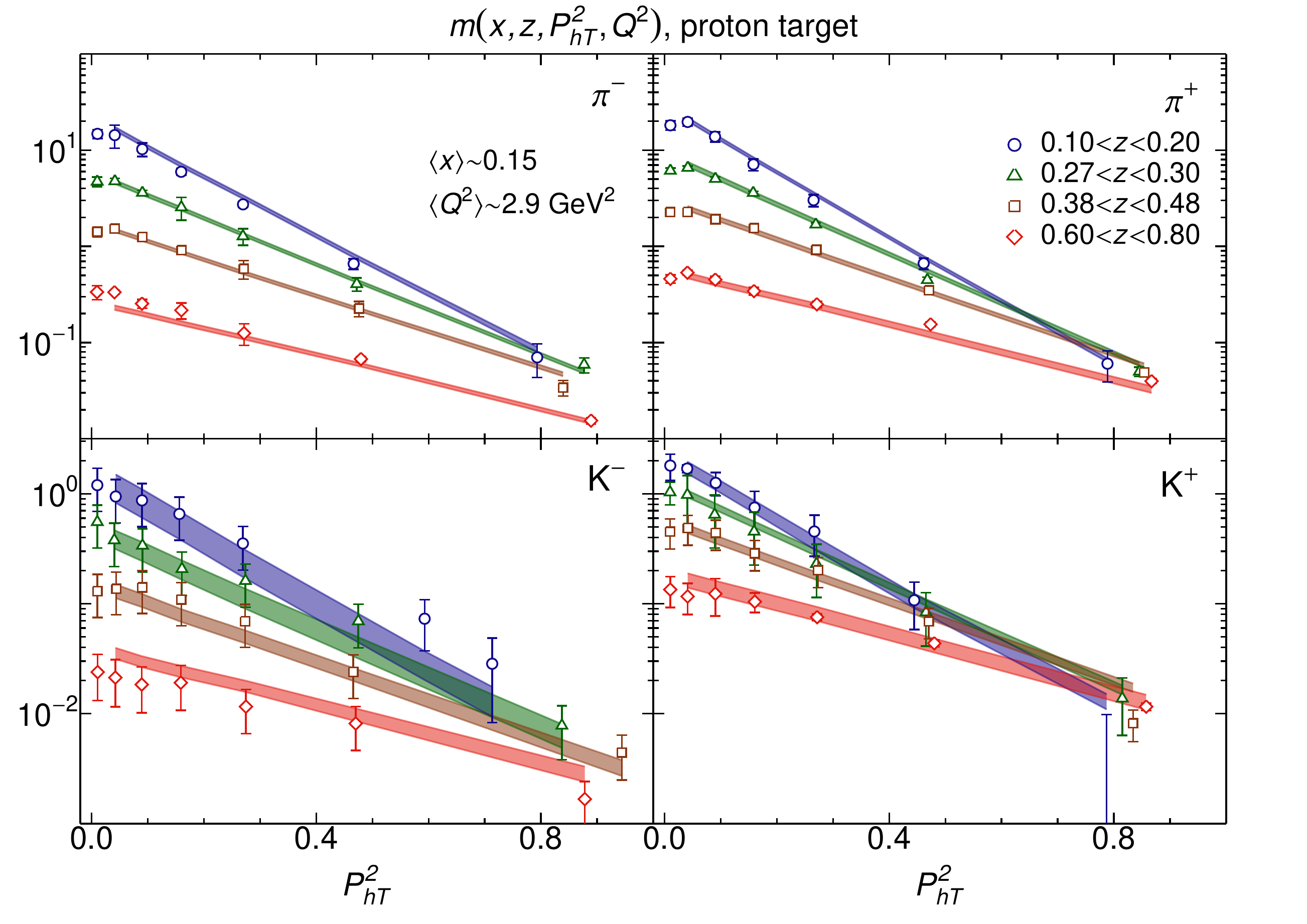}
\caption{Data points: \hermes multiplicities $m_p^h(x,z,\bm{P}_{h\Tperp}^2;
  Q^2)$ for pions and kaons off a proton target as functions of
  $\bm{P}^2_{h\Tperp}$ for one selected $x$ and $Q^2$ bin and few selected $z$
  bins. Shaded bands: 68\% confidence intervals obtained from fitting 200
  replicas of the original data points in the scenario of the default fit. The
  bands include also the uncertainty on the collinear fragmentation
  functions. The lowest $\bm{P}^2_{h\Tperp}$ bin has not been included in the
  fit.} 
\label{f:fit_proton_default}
\end{figure}

\begin{figure}
\centering
\includegraphics[width=13cm]{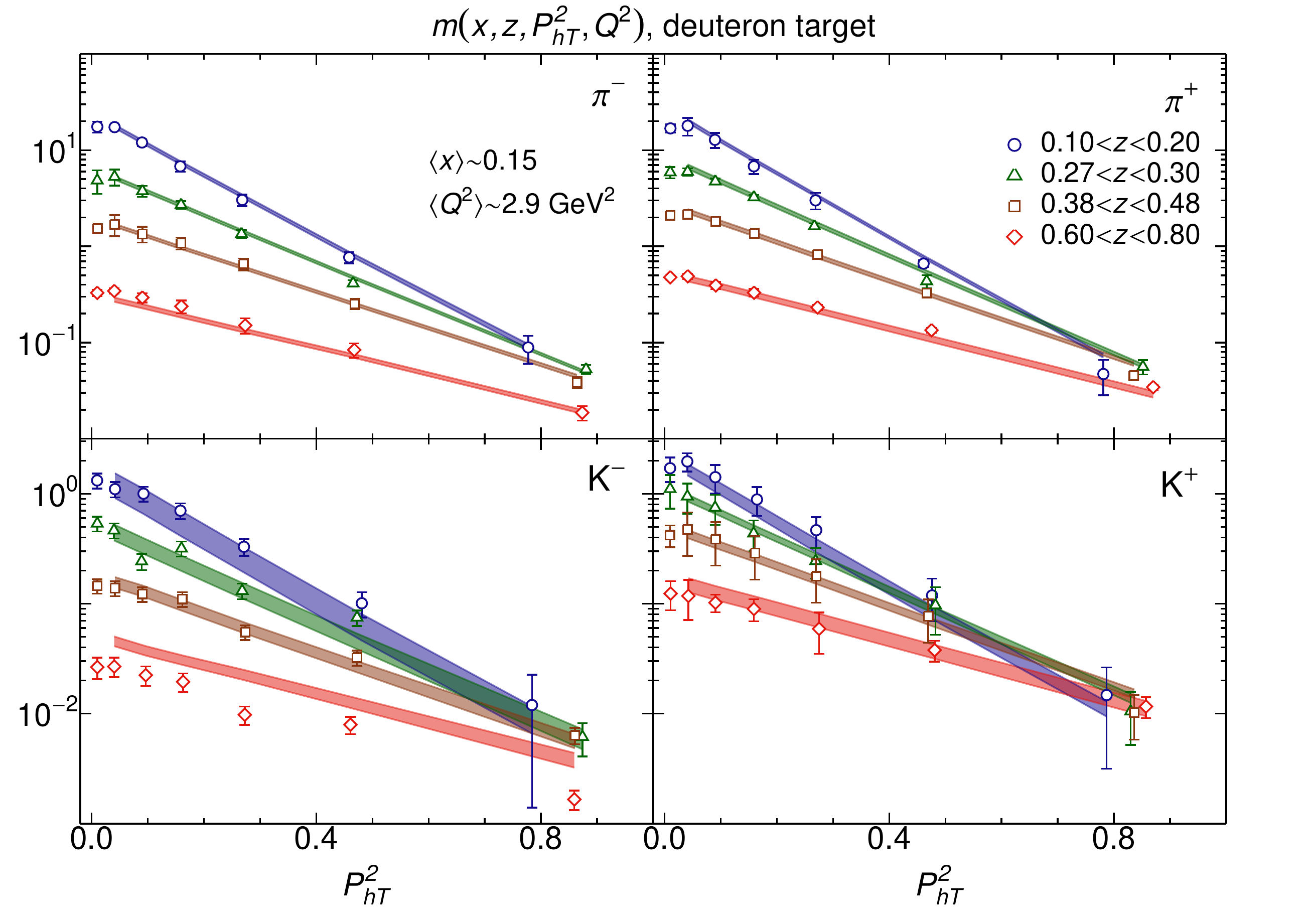}
\caption{Same content and notation as in the previous figure, but for a deuteron target. }
\label{f:fit_deuteron_default}
\end{figure}

Figs.~\ref{f:fit_proton_default} and \ref{f:fit_deuteron_default} illustrate
the agreement between our fit and the \hermes data. For each figure, the upper
panels display the results for pions ($\pi^-$ on the left and $\pi^+$ on the
right), the lower panels for kaons. The results show the multiplicities
$m_N^h(x,z,\bm{P}_{h\Tperp}^2, Q^2)$ for $N=p$ proton and $N=D$ deuteron
targets, respectively, as functions of $\bm{P}^2_{h\Tperp}$ for one selected bin
$\langle x \rangle \sim 0.15$ and $\langle Q^2 \rangle \sim 2.9$ GeV$^2$
(out of the total five $x$ bins we used), and
for four different $z$ bins (out of the total seven $z$ bins we used). 
The lowest $\bm{P}^2_{h\Tperp}$ bin was
excluded from the fit, as explained in  
Sec.\ref{s:selection}. The theoretical band is obtained by rejecting the
largest and lowest $16\%$ of the replicas for each  
$\bm{P}^2_{h\Tperp}$ bin. The theoretical uncertainty is dominated by the
error on the collinear fragmentation functions $D_1(z)$, which induces an
overall normalization uncertainty in each $z$ bin. The different values of the
fit parameters in each replica are 
responsible for the slight differences in the slopes of the upper and lower
borders of the bands. 

In Tab.~\ref{t:fd_PDFs_par}, the values of the average square transverse momenta for TMD PDFs are listed. We note that they 
can range between 0.13 and 0.57 GeV$^2$ within the 68\% confidence interval.

\begin{figure}
\centering
\begin{tabular}{ccc}
\includegraphics[width=8cm]{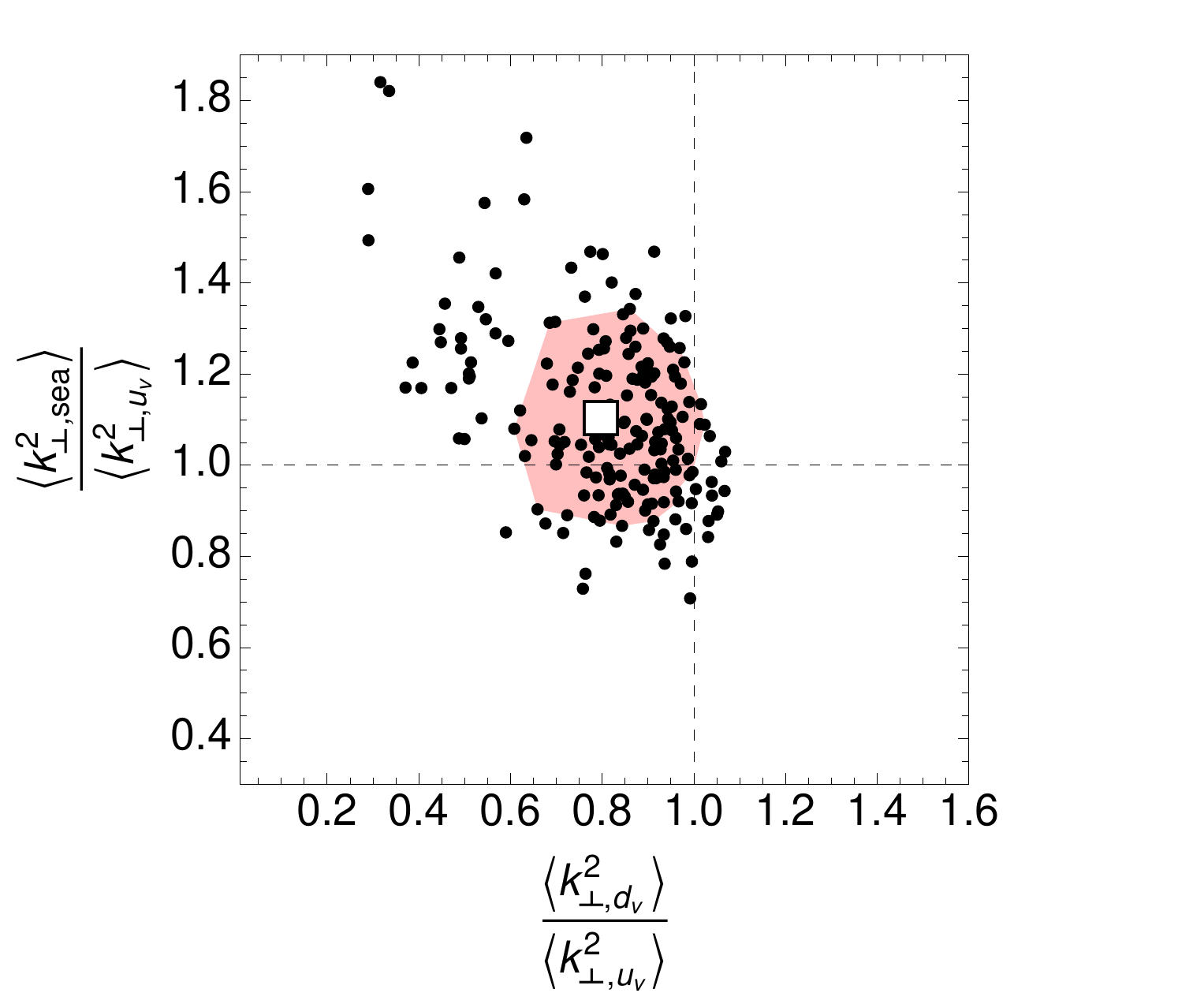}
&\hspace{1cm}
&
\includegraphics[width=8cm]{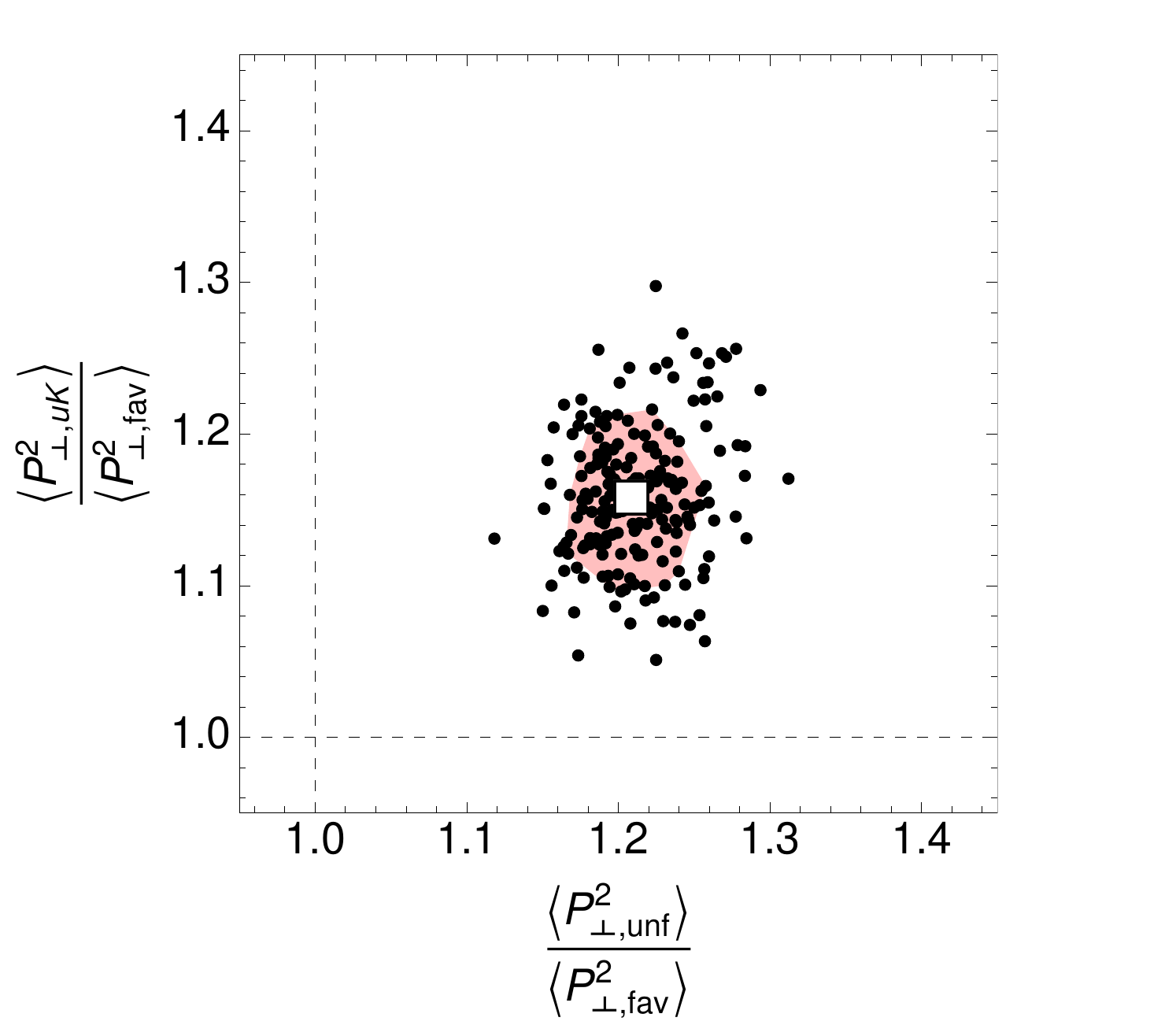}
\\
(a) && (b)
\end{tabular}
\caption{(a) Distribution of the values of the ratios 
$\langle \bm{k}^2_{\T, d_v} \rangle / \langle \bm{k}^2_{\T, u_v}\rangle$ vs. 
$\langle \bm{k}^2_{\T, {\rm sea}}\rangle / \langle \bm{k}^2_{\T, u_v}\rangle$ obtained
from fitting 200 replicas of the original data points in the scenario of the
default fit. The white squared box indicates the center of the 68\% confidence
interval for each ratio.
The shaded area represents the two-dimensional 68\% confidence
region around the white box. 
The dashed lines correspond to the ratios being unity; their crossing
point corresponds to the result with no flavor dependence. For most of the
points,  
$\langle \bm{k}^2_{\T, d_v}\rangle < \langle \bm{k}^2_{\T, u_v}\rangle < \langle
\bm{k}^2_{\T, {\rm sea}}\rangle$.  
(b) Same as previous panel, but for the distribution of the values of the
ratios  $\langle \bm{P}^2_{\T, {\rm
    unf}}\rangle / \langle \bm{P}^2_{\T, {\rm fav}}\rangle$  vs. 
$\langle \bm{P}^2_{\T, u  K}\rangle / \langle \bm{P}^2_{\T, {\rm
    fav}}\rangle$. 
For all points, 
$\langle \bm{P}^2_{\T, {\rm fav}}\rangle < \langle \bm{P}^2_{\T, {\rm
    unf}}\rangle \sim \langle \bm{P}^2_{\T, u K}\rangle$. 
}
\label{f:DoverU_SoverU_default}
\end{figure}

In the left panel of Fig.~\ref{f:DoverU_SoverU_default}, we compare the ratio
$\langle \bm{k}_{\T, d_v}^2 \rangle / \langle \bm{k}_{\T, u_v}^2 \rangle$
vs. $\langle \bm{k}_{\T, {\rm sea}}^2 \rangle / \langle \bm{k}_{\T, u_v}^2
\rangle$ for 200 replicas. The white box represents the point at
the center of each one-dimensional 68\% confidence interval of the two
ratios. 
The shaded area represents the two-dimensional 68\%
confidence region, it contains 68\% of the points with the shortest
distance from the white box. 
Since for each
flavor the $x$ dependence of the average square transverse momenta is the same
(see Eq.~\eqref{e:kT2_kin}), these ratios are $x$-independent. The dashed
lines 
correspond to the ratios being unity and divide the plane into four
quadrants. Most of the replicas are in the upper left quadrant, i.e.,   we
have 
$\langle \bm{k}_{\T, d_v}^2 \rangle < \langle \bm{k}_{\T, u_v}^2 \rangle < \langle \bm{k}_{\T, {\rm sea}}^2 \rangle$.
%The average values of these ratios (denoted by the white box in the figure)
The white box shows that $d_v$ is on average about 20\% narrower than  $u_v$, which is in
turn about 10\% narrower than the sea. The crossing of the dashed lines
corresponds to a flavor-independent distribution of transverse momenta. 
This crossing point lies at the limit of the 68\% confidence region. In a
relevant number of replicas $d_v$ can be more than 40\% narrower than the
$u_v$, and the sea can be more than 30\% wider than $u_v$. From this fit, it
seems possible that the sea is narrower than
$u_v$, but unlikely that $d_v$ is wider than $u_v$.  

In the right panel of Fig.~\ref{f:DoverU_SoverU_default}, we compare the ratio
$\langle \bm{P}_{\T, {\rm unf}}^2 \rangle / \langle \bm{P}_{\T, {\rm fav}}^2
\rangle$ vs. $\langle \bm{P}_{\T, u K}^2 \rangle / \langle \bm{P}_{\T, {\rm
    fav}}^2 \rangle$ in the same conditions as before. All points are
clustered in the upper right quadrant and close to its bisectrix, i.e.,  
we have the stable outcome that 
$\langle \bm{P}_{\T, {\rm fav}}^2 \rangle < \langle \bm{P}_{\T, {\rm unf}}^2
\rangle \sim \langle \bm{P}_{\T, u K}^2 \rangle$. The width of unfavored and
$u \to K^+$ fragmentations are about 20\% larger than the widht of favored ones.

%%%%%

%%%%%%%%%%%%%%%%%%%%%%%%
\subsection{Fit with $Q^2>1.6 \text{ GeV}^2$}
\begin{figure}
\centering
\begin{tabular}{ccc}
\includegraphics[width=8cm]{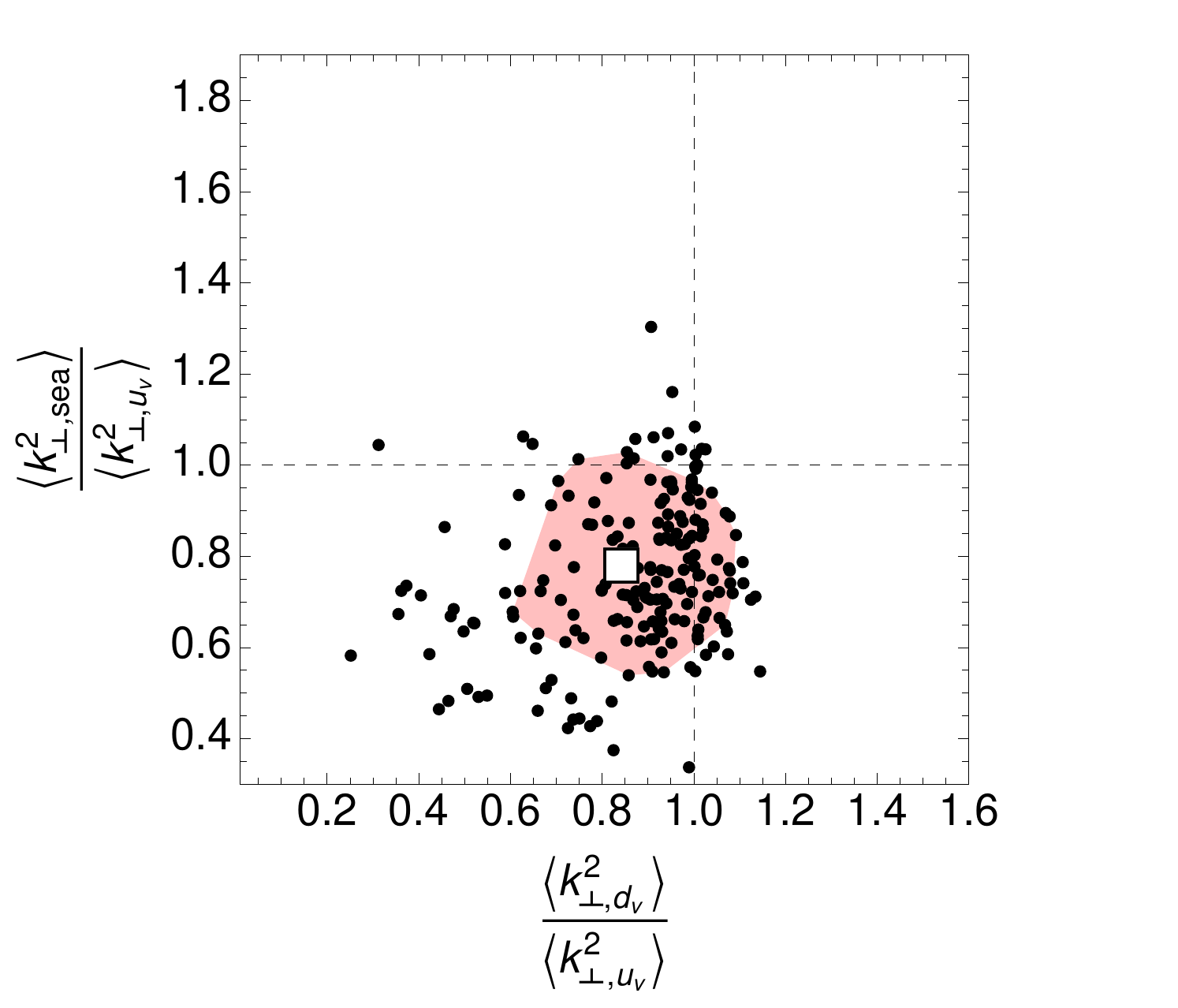}
&\hspace{1cm}
&
\includegraphics[width=8cm]{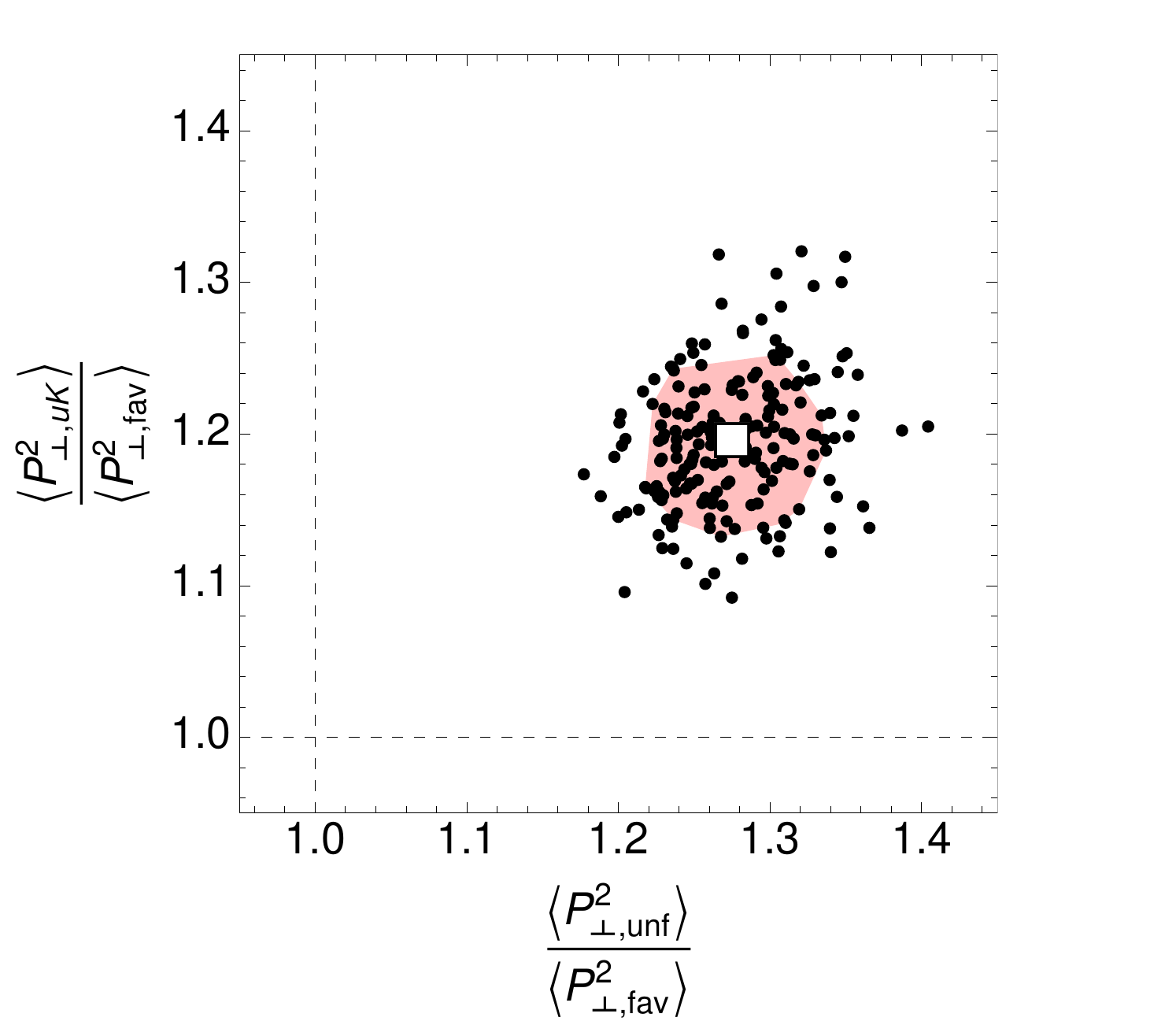}
\\
(a) && (b)
\end{tabular}
\caption{Same content and notation as in the previous figure, but for the scenario with the cut $Q^2 > 1.6$ . }
\label{f:DoverU_SoverU_highQ2}
\end{figure}
In this scenario, we restrict the $Q^2$ range compared to the default fit by imposing the cut $Q^2>1.6$ GeV$^2$. The set of data is reduced to 1274 points.
%% chi20d0f
The mean value of the $\chi^2$/d.o.f is smaller, since we are fitting less data. Moreover, the disregarded $Q^2$ bin contains high statistics.
As for the default fit, the behavior of transverse momenta over the 200 replicas is summarized in Fig. \ref{f:DoverU_SoverU_highQ2}.
%% sea quarks smaller
The exclusion of low-$Q^2$ data leads to partial differences in the features
of the extracted TMD PDFs: the
average width of valence quarks slightly increases, while the distribution for
sea quarks becomes narrower. 
%This behavior is not surprising: at
%low $Q^2$, hence at low $x$, sea quarks are dominant; at larger $x$, the
%density of sea quarks with nonnegligible transverse momenta reduces. 
In the
left panel, most of the replicas are in the lower left quadrant, i.e.,   we
have $\langle \bm{k}_{\T, {\rm sea}}^2 \rangle \lesssim \langle \bm{k}_{\T,
  d_v}^2 \rangle < \langle \bm{k}_{\T, u_v}^2 \rangle$. 
On average, $d_v$ quarks are 15\% narrower than than $u_v$ quarks, which are
in turn more than 20\% wider than sea quarks. 
In a relevant number of replicas $d_v$ can be more than 40\% narrower than the $u_v$, and the sea can be even 50\% narrower than $u_v$.
In this scenario, it is unlikely that the sea is wider than $u_v$, but it is
possible that $d_v$ is wider than $u_v$. 

In the right panel, the behavior of transverse momenta in fragmentation
processes is qualitatively unchanged with respect to the default fit, apart
from the fact that the unfavored Gaussian function becomes now more than 25\%
larger than the favored one. 

The crossing point again indicates no flavor dependence and lies just outside
the 68\% confidence region for TMD PDFs and completely outside the same region
for TMD FFs.  

We conclude that the low-$Q^2$ data, being also characterized by low $x$, can
have a significant impact on the analysis of TMD PDFs, in particular the sea
components. More data at low $x$ (but possibly at high $Q^2$) 
are necessary to better constrain the sea quarks TMD
PDFs~\cite{Boer:2011fh,Accardi:2012hwp,Adolph:2013stb}.  

%%%%%%%

\subsection{Fit with pions only}
We also choose to fit data related only to pions in the final state, in order to explore the importance of the kaons data set.
In this framework, we are left with two independent fragmentation processes: favored and unfavored ones.
Accordingly, the number of fit parameters for TMD FFs reduces from 7 to 5 ($\langle \hat{\bm{P}}^2_{\T, {\rm fav}}\rangle$, 
$\langle \hat{\bm{P}}^2_{\T, {\rm unf}}\rangle$, $\beta$, $\delta$, $\gamma$; see 
Eqs.~\eqref{e:favored}-\eqref{e:PT2_kin}).
\begin{figure}
\centering
\begin{tabular}{ccc}
\includegraphics[width=8cm]{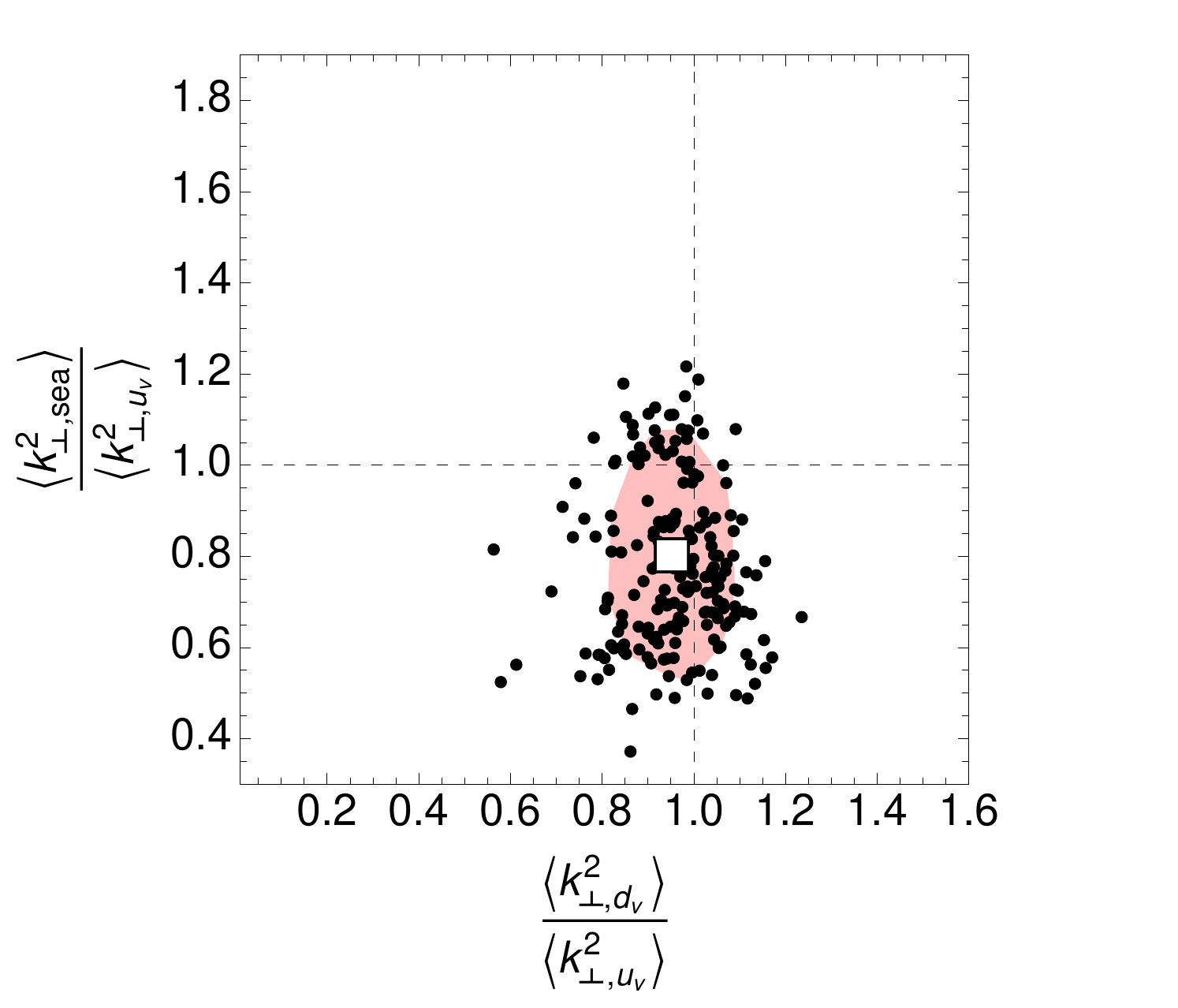}
&\hspace{1cm}
&
\includegraphics[width=8cm]{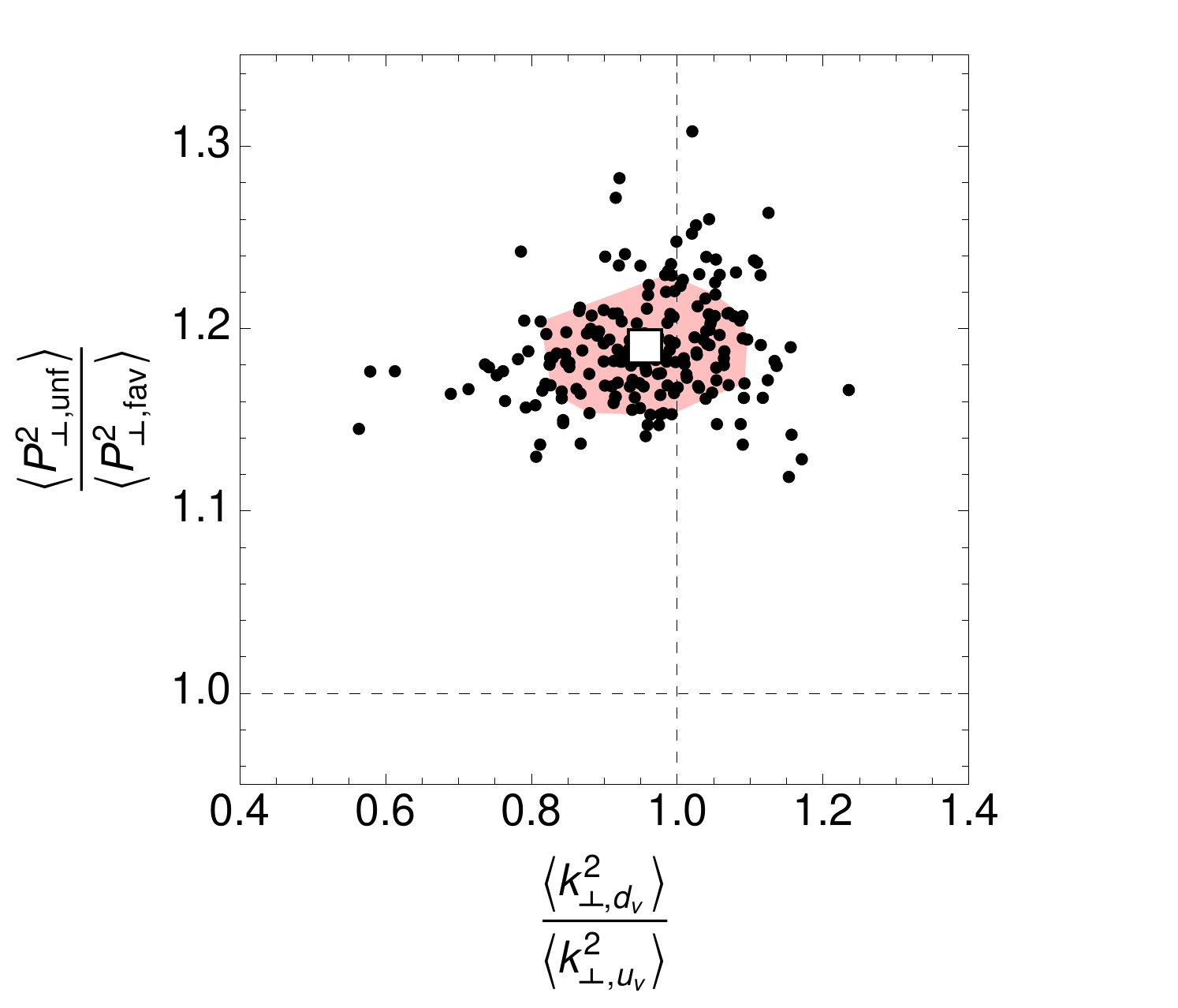}
\\
(a) && (b)
\end{tabular}
\caption{(a) Same content and notation as in Fig.~\ref{f:DoverU_SoverU_default}a) but for the scenario with only pions in the final state. For most of the points, $\langle \bm{k}^2_{\T,{\rm sea}} \rangle < \langle \bm{k}^2_{\T, d_v}\rangle \lesssim \langle \bm{k}^2_{\T, u_v}\rangle$. (b) Distribution of the values of the ratios 
$\langle \bm{P}^2_{\T, {\rm unf}}\rangle / \langle \bm{P}^2_{\T, {\rm fav}}\rangle$ vs. 
$\langle \bm{k}^2_{\T, d_v}\rangle / \langle \bm{k}^2_{\T, {u_v}}\rangle$ obtained in the same conditions as in the previous case. For all points $\langle \bm{P}^2_{\T, {\rm fav}}\rangle < \langle \bm{P}^2_{\T, {\rm unf}}\rangle$. }
\label{f:DoverU_SoverU_pions_only}
\end{figure}
%
%{\bf Recall that in this case we have only two channels for TMD FFs: favored and unfavored}

The agreement between data and the model is the worst (see
Tab. (\ref{t:chi2collinear})). This is due to at least two reasons. First of
all, the fit of collinear multiplicities was poor in all the target-hadron
combinations involving pions in the final state. Moreover, the high statistics
collected for pions (mostly in the low-$Q^2$ region) leads to higher values of
$\chi^2$.  

% TMD PDFs
Fig.~\ref{f:DoverU_SoverU_pions_only} shows the behavior of transverse momenta over the 200 replicas. As for TMD PDFs, in the left panel most of the replicas are in the lower part, i.e.,  we have $\langle \bm{k}_{\T, {\rm sea}}^2 \rangle < \langle \bm{k}_{\T, d_v}^2 \rangle \lesssim \langle \bm{k}_{\T, u_v}^2 \rangle$.
On average, $d_v$ quarks are equally distributed as $u_v$ quarks, which are in
turn more than 20\% wider than sea quarks.
In the default fit sea quarks were wider than valence ones and there was a
remarkable difference between $u_v$ and $d_v$, not evident in this scenario. 
In any case, 
in a relevant number of replicas $d_v$ can be more than 15\% narrower than the
$u_v$, but also more than 10\% wider than $u_v$. 
The sea can be even 50\% narrower than $u_v$, but it is also 
not unlikely that the sea is wider than $u_v$
Once again, 
the crossing point for flavor independence lies at the boundary of the 68\%
confidence region, due to the difference between the
distributions of sea quarks and valence quarks.

% TMD FFs 
As in the other scenarios, TMD FFs for unfavored processes are wider than
favored ones. 
The difference is comparable to the default fit, with unfavored functions
about 20\% larger than favored.

Similar fits have been performed in
Refs.~\cite{Mkrtchyan:2007sr,Asaturyan:2011mq}, but using data averaged over
$z$, which renders it particularly difficult to disentangle the distribution
and fragmentation contributions. To overcome this problem, 
both fits included also indirect information from the
azimuthal $\cos \phi_h$ dependence. 
The fit of Ref.~\cite{Asaturyan:2011mq} obtained 
a small value for the
distribution mean square transverse momenta of up quarks, $\langle \bm{k}_{\T,u}^2
\rangle = 0.07\pm 0.03 \text{ GeV}^2$, while the down quark mean transverse
momentum was compatible with zero, $\langle \bm{k}_{\T,u}^2
\rangle = -0.01\pm 0.05 \text{ GeV}^2$ (sea quarks were
neglected).
The previous
fit~\cite{Mkrtchyan:2007sr}, obtained a somewhat different behavior, with 
a mean transverse momentum of the up quark compatible with zero and 
$\langle \bm{k}_{\T,d}^2 \rangle = 0.11\pm 0.13 \text{ GeV}^2$. 
In both fits, the average values of the width of the TMD FFs are compatible
with our results, but, contrary to our findings, a slight tendency for the
favored FF to be larger than unfavored was found.
In any case, we remark that the average kinematics of the experiment taken
into consideration in Refs.~\cite{Mkrtchyan:2007sr,Asaturyan:2011mq} are
different from \hermes
 (see also the discussion in
 Ref.~\cite{Schweitzer:2010tt}). 

Overall, we conclude that kaon data have an important impact in a
flavor-dependent analysis, due to the large 
role played by strange quarks and
antiquarks in kaon multiplicities.

%%%%%%%%%%%%%%

\subsection{Flavor-independent fit}
\begin{figure}
\includegraphics[width=8cm]{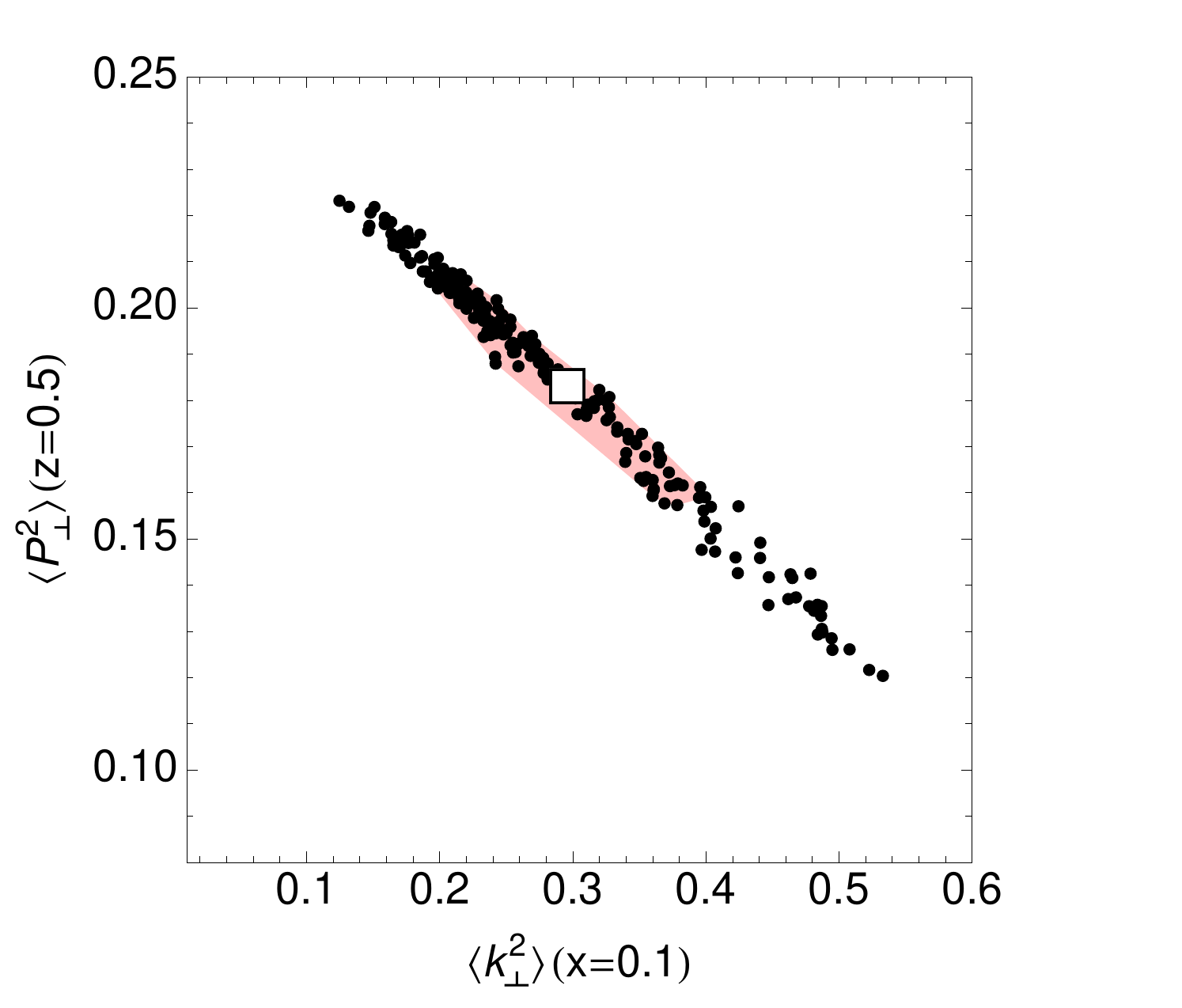}
\caption{Distribution of the values of $\langle \bm{k}^2_{\T} \rangle$ (at $x=0.1$) and
  $\langle \bm{P}^2_{\T}\rangle$ (at $z=0.5$) obtained from fitting 200 replicas of the
  original data points in the scenario of the flavor-independent fit. The
  white squared box indicates the center of the one-dimensional 68\%
  confidence interval for each parameter. The shaded
  area represents the two-dimensional 68\% confidence region around the white
  box. 
The transverse
  momenta are manifestly anti-correlated. } 
\label{f:fi_kT2vsPT2}
\end{figure}
In this scenario, we assume a Gaussian ansatz for unpolarized TMD PDFs and TMD FFs with flavor-independent widths, 
i.e.,  we neglect any flavor dependence in Eq.~\eqref{e:fldep_gauss}: 
\begin{equation}
\big \langle \hat{\bm{k}}_{\T,u_v}^2 \big \rangle = \big \langle \hat{\bm{k}}_{\T,d_v}^2 \big \rangle = \big \langle \hat{\bm{k}}_{\T,{\rm sea}}^2 \big \rangle \equiv \big \langle \hat{\bm{k}}_\T^2 \big \rangle  \ ,
\label{e:fi_PDFs}
\end{equation}
\begin{equation}
\big \langle \hat{\bm{P}}_{\T,{\rm fav}}^2 \big \rangle  = \big \langle
\hat{\bm{P}}_{\T,{\rm unf}}^2 \big \rangle  = \big \langle \hat{\bm{P}}_{\T,u  K}^2 \big \rangle  = \big \langle \hat{\bm{P}}_{\T,s  K}^2 \big \rangle  \equiv \big \langle \hat{\bm{P}}_\T^2 \big \rangle  \ .
\label{e:fi_FFs}
\end{equation}
Accordingly, the number of fit parameters reduces to 3 for TMD PDFs ($\langle
\hat{\bm{k}}_\T^2 \rangle$, $\alpha$, $\sigma$) 
and 4 for TMD FFs ($\langle \hat{\bm{P}}_\T^2
\rangle$, $\beta$, $\delta$, $\gamma$). 
Their values are summarized in Tab. \ref{t:fd_PDFs_par}
and \ref{t:fd_FFs_par}. 
The expression~\eqref{e:FDmult} for the multiplicities considerably simplifies and the ${\bm P}_{h\Tperp}$ width is the same for every target-hadron combination:
\begin{equation}
\big \langle \bm{P}_{h\Tperp}^2 \big \rangle = z^2 \big \langle \bm{k}_{\T}^2 \big \rangle + \big \langle \bm{P}_{\T}^2 \big \rangle \ .
\label{e:fi_TMrel}
\end{equation}

The agreement between data and the flavor-independent model is poorer than in
the (flavor-dependent) default fit: the central value of the $\chi^2$/d.o.f. is
$1.73$ (see Tab. \ref{t:fd_chi20dof}). 
This is not surprising, since we are fitting with the same function data for
all the available target-hadron combinations, which display sensibly different
behaviors. 
However, these results do not rule out the flavor-independent ansatz. 

Fig.~\ref{f:fi_kT2vsPT2} clearly shows the anti-correlation between $\langle
\hat{\bm{k}}_{\T}^2 \rangle$ and $\langle \hat{\bm{P}}_{\T}^2 \rangle$ induced
by Eq.~\eqref{e:fi_TMrel}. 

Similar fits have been performed in the past for semi-inclusive DIS and Drell--Yan
processes~\cite{D'Alesio:2004up,Schweitzer:2010tt}, also including the effect
of gluon resummation~\cite{Nadolsky:2000ky,Landry:2002ix,Konychev:2005iy}
within the so-called 
Collins--Soper--Sterman formalism~\cite{Collins:1984kg,Bozzi:2010xn}, which is
equivalent to taking into account TMD
evolution~\cite{Aybat:2011zv,Collins:2011zzd,Echevarria:2012pw}. 

The values of our mean square transverse momenta at $x=0.1$ and $z=0.5$ are
consistent with the values obtained without considering $x$ and $z$ dependence
in Ref.~\cite{Schweitzer:2010tt}
($\langle \bm{k}_\T^2 \big \rangle = 0.38 \pm 0.06 \text{  GeV}^2$  and 
$\langle \bm{P}_\T^2 \big \rangle = 0.16 \pm 0.01 \text{  GeV}^2$)
and in Ref.~\cite{Anselmino:2005nn} 
($\langle \bm{k}_\T^2 \big \rangle = 0.25 \text{  GeV}^2$  and 
$\langle \bm{P}_\T^2 \big \rangle = 0.20 \text{  GeV}^2$)
using a different approach based on the
so-called Cahn effect~\cite{Cahn:1978se}. 
In the \hermes  Monte Carlo generator {\sc GMC$_{\text{TRANS}}$}, the following
flavor-independent parametrization of the mean square transverse momenta, which were tuned to \hermes pion-multiplicity data,  has
been implemented:
\begin{align}  
 \langle \bm{k}_\T^2 \rangle &= 0.14 \text{ GeV}^2 ,
&
\langle \bm{P}_\T^2 \rangle &= 0.42 \, z^{0.54} (1-z)^{0.37} \text{ GeV}^2 .
\label{e:avpT}
\end{align} 
The latter functional form is not much different from the one we obtained. The
value of the distribution transverse momentum is slightly smaller than
our average value, which is compensated by the fact that the fragmentation
transverse momentum is slightly higher. Other fits that explored the $z$
dependence in the Gaussian width of TMD FFs can be found in 
Refs.~\cite{Boglione:1999pz,D'Alesio:2004up}.

Comparison with extractions from Drell--Yan experiments (see, e.g., 
Refs.~\cite{D'Alesio:2004up,Schweitzer:2010tt,%
Nadolsky:2000ky,Landry:2002ix,Konychev:2005iy})  
is not straightforward, due 
to the different kinematic conditions and the difficulty to extrapolate the
results obtained in the CSS formalism  (see also the discussion in
Ref.~\cite{Sun:2013hua}). The mean square transverse
momentum obtained from Gaussian fits without TMD evolution
\cite{D'Alesio:2004up,Schweitzer:2010tt}
is larger than in our case, 
$\langle \bm{k}_\T^2 \big \rangle \gtrsim 0.7 \text{  GeV}^2$.

%%%%%%%%%%%%%%%%%%%%%%%%%%%%%%%%%%%%%%%%%%%%
\section{Conclusions}
%%%%%%%%%%%%%%%%%%%%%%%%%%%%%%%%%%%%%%%%%%%%
\label{s:conclusion}
%% summary of the procedure

Using the recently published \hermes  data on semi-inclusive DIS
multiplicities~\cite{Airapetian:2012ki}, we explored for the first
time the flavor dependence of the transverse momenta of both the unpolarized
parton distributions (TMD PDFs) and fragmentation functions (TMD FFs). We adopted
a simplified framework based on the parton model and neglecting the effects
of QCD evolution. Using a flavor-dependent Gaussian ansatz, we obtained different
results for multiplicities in eight different target--hadron
combinations. 
We
performed several fits of the data in different scenarios: 
including all bins as described in Sec.~\ref{s:selection}
(the ``default fit"), excluding data with \( Q^{2} \le 1.6~\text{GeV}^{2} \)
%the lowest-$Q^2$ bin 
(equivalent to excluding partons at low $x$), 
selecting only pions in the final state, 
or neglecting any flavor dependence.  

%% relation between flavor dependence and independence, F_UU,T not Gaussian
%% any more 
Comparing the default fit and the flavor-independent one, we conclude
that the flavor-dependent Gaussian ansatz performs better. The
difference between the average $\chi^2$/d.o.f. in the two cases is not
striking but, nonetheless, appreciable. We find convincing
indications that the unfavored fragmentation functions have larger average
transverse momenta with respect to pion favored fragmentation functions. 
We get weaker
indications of flavor dependence for the TMD PDFs. It is very likely to find
fits of the available data with differences of the order of
20\% in the the mean square transverse
momenta of different flavors. 
In particular, our default fit shows a tendency 
for valence down quarks to have a narrower distribution than the one of valence up quarks, 
which in turn is narrower than the one for sea quarks. 
%to have
%$\langle \bm{k}_{\T, d_v}^2 \rangle < \langle \bm{k}_{\T, u_v}^2 \rangle < \langle \bm{k}_{\T, {\rm sea}}^2 \rangle$. 
These features have a potentially large impact on the polarization-dependent
TMD extractions~\cite{Anselmino:2012aa,Bacchetta:2011gx,Anselmino:2008sga,%
Arnold:2008ap,Barone:2009hw,Lu:2009ip,Anselmino:2008jk}, 
where usual flavor-independent 
transverse momentum parametrizations are assumed in the fragmentation, 
as even the normalizations extracted for those TMDs 
depend, directly or indirectly, on the widths of the polarization-averaged TMD
FFs.

Apart from the ratios among different flavors, the absolute values of the mean
square transverse momenta are compatible with results quoted in the
literature. However, it should be kept in mind that there exist   
strong anti-correlations
between mean squared transverse momenta of distribution and fragmentation functions.

This work is a first step in the exploration of the transverse momentum
dependence of partons inside hadrons. First of all, it needs to be updated by
implementing evolution equations in the TMD
framework~\cite{Aybat:2011zv,GarciaEchevarria:2011rb,Echevarria:2012pw,%
Collins:2012uy,Sun:2013hua}. Secondly,
the data set needs to be enlarged to include the recently released \compass
data in a wider kinematical domain~\cite{Adolph:2013stb}, and, in the
following step, to include also data from $e^+ e^-$ annihilations and
Drell-Yan processes. Finally, other functional forms different from the
Gaussian ansatz should be explored.

%%%%%%%%%%%%%%%%%%%%%%%%%%%%%%%%%%%%%%%%%%%%
\acknowledgments
%%%%%%%%%%%%%%%%%%%%%%%%%%%%%%%%%%%%%%%%%%%%
Discussions with Maarten Buffing, Marco Contalbrigo, Jasone Garay Garc\'ia,
Marco Guagnelli, Piet J.~Mulders, Barbara Pasquini, Jean-Francois Rajotte, 
Marco Stratmann, and Charlotte Van Hulse are gratefully acknowledged. The work
of A.~S. is part of the program of the ``Stichting voor Fundamenteel Onderzoek
der Materie'' (FOM), which is financially supported by the Nederlandse
Organisatie voor Wetenschappelijk Onderzoek (NWO).  
This work is partially supported by the European Community through the
Research Infrastructure Integrating Activity ``HadronPhysics3'' (Grant
Agreement n. 283286) under the European 7th Framework Programme, 
the Basque Foundation for Science (IKERBASQUE) and the UPV/EHU under program UFI 11/55.

%%%%%%%%%%%%%%%%%%%%%%%%%%%%%%%%%%%%%%%%%%%%%%%%%%%%%%%%%%%%%

\bibliography{mybiblio}

%%%%%%%%%%%%%%%%%%%%%%%%%%%%%%%%%%%%%%%%%%%%%%%%%%%%%%%%%%%%%

\end{document}